\documentclass[prb,amssymb,twocolumn,showpacs,supersriptaddress,floatfix]{revtex4}
\usepackage{amssymb}
\usepackage{amsmath}
\usepackage{graphicx,subfigure}
\usepackage{bm}

\begin{document}

\title{Dynamic transition  in Landau-Zener-St\"uckelberg interferometry of dissipative systems: the case of the flux qubit}

\author{Alejandro Ferr\'on}
\affiliation{Instituto de Modelado e Innovaci\'on Tecnol\'ogica (CONICET-UNNE), 3400 Corrientes, Argentina.}
\author{Daniel Dom\'{\i}nguez}
\author{Mar\'{\i}a Jos\'{e} S\'{a}nchez}
\affiliation{Centro At{\'{o}}mico Bariloche and Instituto Balseiro, 8400 San Carlos de Bariloche, R\'{\i}o Negro, Argentina.}

\begin{abstract}
	
We study  Landau-Zener-Stuckelberg (LZS) interferometry in multilevel systems coupled to an Ohmic quantum bath.
We consider the case of superconducting  flux qubits  driven  by a dc+ac magnetic fields, but our results can apply to other similar systems. 
We find a dynamic transition  manifested by a symmetry change in the structure of the LZS interference pattern, plotted as a function of ac amplitude and dc detuning.
The dynamic transition is from a LZS pattern with nearly symmetric multiphoton resonances to antisymmetric multiphoton 
resonances at long times (above the relaxation time).  
We also show that the presence of a resonant mode  in the quantum bath can
impede the  dynamic transition when the resonant frequency is
of the order of the qubit gap.
Our results are obtained by a numerical calculation of the finite time and the asymptotic stationary population of the qubit states, using the Floquet-Markov approach to
solve a realistic model of the flux qubit considering up to 10 energy levels. 

\end{abstract}


\maketitle

\section{Introduction}

In recent years, the experimental realization 
of Landau-Zener-St\"{u}ckelberg (LZS) interferometry\cite{shevchenko}  
in several systems has emerged as 
a tool to study quantum coherence under strong driving.
LZS interferometry is realized in two-level systems (TLS)
which are  driven with a time periodic force.
The periodic sweeps through an avoided crossing in the energy level spectrum
result in successive Landau-Zener transitions.
The accumulated phase between these
repeated tunneling events gives place to 
constructive or destructive interferences,
depending on the driving amplitude and the detuning 
from the avoided crossing.
These LZS interferences have been observed in a variety of
quantum systems, such as Rydberg atoms,\cite{yoakum} superconducting qubits,\cite{oliver,izmalkov,berns1,cqubits,rudner,valenzuela,wilson,wilson2,sun,sun2,wang2010,graaf2013,shevchenko2}
 ultracold molecular gases,\cite{mark}
 quantum dot devices,\cite{petta2010,stehlik2012,cao2013,dupont2013,shang2013,nalbach2013,ludwig2014,stace2014,granger2015}
 single spins in  nitrogen vacancy centers in diamond,\cite{huang,zhou2014}  nanomechanical circuits,\cite{lahaye} and ultracold atoms in accelerated optical lattices.\cite{kling} 
  Several other related experimental and  theoretical 
  works have studied LZS interferometry in systems 
  under different shapes  of periodic driving,\cite{bylander,gustavsson,satanin2014,blattmann,xu,silveri2015}
  in two coupled qubits,\cite{satanin2012} in optomechanical systems,\cite{marquardt2010} 
  and the effect of a geometric phase.\cite{geometric}
 Furthermore, experiments in superconducting flux qubits under strong driving
 have allowed to
extend LZS interferometry  beyond two levels.\cite{valenzuela}
In this later case, the multi-level structure of the flux qubit,  
with several different avoided crossings in the energy spectrum,
exhibited a series of diamond-like interference patterns 
as a function of
dc flux detuning and microwave amplitude.\cite{valenzuela,wen}

Driven two-level systems have been extensively studied theoretically in the past. 
Under strong time-periodic driving fields, 
phenomena such as coherent destruction of tunneling (CDT)
\cite{grossmann,kayanuma} and multiphoton resonances \cite{shirley,ashhab}
have been analyzed. The influence of the environment has been studied
within the driven spin-boson model,\cite{grifoni-hanggi} applying various techniques
like the path-integral formalism,\cite{hartmann}  or the solution of
the time dependent equations for the populations 
of the density matrix,\cite{grifoni-hanggi,hartmann,popin2l,goorden2003,stace2005,grifonih}
either as an integro-differential kinetic equation,\cite{popin2l} or considering
for weak coupling the underlying Bloch-Redfield
equations,\cite{hartmann,goorden2003}
or using the decomposition of the quantum master equation into Floquet states,\cite{grifoni-hanggi,grifonih}
 or performing a rotating-wave approximation.\cite{stace2005}
However, the   theoretical results for  driven TLS in contact with a quantum bath \cite{grifoni-hanggi,hartmann,popin2l,goorden2003,stace2005,grifonih} 
present some discrepancies with  the experiments of LZS interferometry, particularly in the flux qubit.\cite{oliver,berns1,rudner,valenzuela} In fact, these works typically predict population inversion,\cite{hartmann,popin2l,goorden2003,stace2005,grifonih}
which has  not been observed in the experiments of Ref.~\onlinecite{oliver,berns1,rudner,valenzuela}.
In a recent work we showed,\cite{fds2} that for the case of a non-resonant detuning, 
population inversion arises for very long
driving times and it is mediated by a slow mechanism of interactions with the bath.  
This long-time asymptotic regime was not  reached in the experiment.

Here, we will study in full detail the LZS interference patterns of the flux qubit,
by considering the dependence with dc flux detuning in addition to the dependence with microwave amplitude. 
To  this end, we calculate numerically the finite time and the asymptotic stationary population of the qubit states using the Floquet-Markov approach,\cite{grifoni-hanggi,fds2} solving a realistic model of the flux qubit  in contact with an Ohmic quantum bath. 
For the time scale of the experiments  we  find a very good 
agreement with the diamond patterns of Ref.\onlinecite{valenzuela}. 
For longer time scales, we find a dynamic transition within the first diamond (which corresponds to the
two level regime), manifested by a symmetry change in the structure
of the LZS interference pattern. We also  consider the case
of an structured quantum bath, which can be due to a SQUID detector with a resonant plasma frequency $\Omega_p$. Different
types of LZS interference patterns can arise in this case, depending
on the magnitude of $\Omega_p$.

The paper is organized as follows. In Sec.II we present the model
for the flux qubit and we review the basics of LZS interferometry.
In Sec. III, we show our results for the emergence of a dynamic transition in the LZS
interference pattern at long times.
In Sec. IV we give our conclusions and an Appendix is included
to describe in detail  the Floquet-Markov formalism used in this paper.

\section{Review of Landau-Zener-Stuckelberg interferometry of the Flux Qubit}
\label{model}
\subsection{The flux qubit}

Superconducting circuits with  mesoscopic Josephson junctions
are used as quantum bits\cite{revqubits}
and can behave as artificial atoms.\cite{nori2011}. 
A well studied  circuit is the
flux qubit (FQ) \cite{qbit_mooij,chiorescu,chiorescu2}
which, for millikelvin temperatures, exhibits
quantized energies levels  that are sensitive to  an external magnetic field. 
The FQ consists on a superconducting ring with  three Josephson
junctions\cite{qbit_mooij} enclosing a magnetic flux $\Phi=
f\Phi_0$, with $\Phi_0=h/2e$. The circuits that are used for the FQ
have two of the junctions with the same Josephson coupling energy,
$E_{J,1}=E_{J,2}=E_J$, and capacitance, $C_1=C_2=C$, while the
third junction has smaller coupling $E_{J,3}=\alpha E_J$ and
capacitance $C_3=\alpha C$, with $0.5<\alpha<1$.
The junctions have gauge invariant phase differences defined as
$\varphi_1$, $\varphi_2$ and $\varphi_3$, respectively. 
Typically the circuit inductance can
be neglected  and the phase difference of the third junction is:
$\varphi_3=-\varphi _1 +\varphi _2 -2\pi f$.  Therefore, the system
can be described with  two independent dynamical variables. A convenient choice
is $\varphi_l=(\varphi_1-\varphi_2)/2$ (longitudinal phase)
$\varphi_t=(\varphi_1+\varphi_2)/2$ (transverse phase).  In terms
of this variables, the hamiltonian of the FQ (in units of $E_J$) is:\cite{qbit_mooij}

\begin{equation}
	\label{ham-sys}
	{\cal H}_{FQ}=-\frac{\eta^2}{4}\left(\frac{\partial^2}{\partial\varphi_t^2}+
	\frac{1}{1+2\alpha}\frac{\partial^2}{\partial\varphi_l^2}\right)
	+V(\varphi_l,\varphi_t)\; ,
\end{equation}
with $\eta^2=8E_C/E_J$ and $E_C=e^2/2C$.
The kinetic term of the hamiltonian corresponds to the electrostatic energy
of the system  and the potential term corresponds to the Josephson energy
of the junctions, given by
\begin{equation}
\label{eq:pot}
V(\varphi_l,\varphi_t)= 2+\alpha -
2\cos\varphi_t\cos \varphi_l - \alpha \cos (2\pi f+2\varphi _l) \;
\end{equation}
Typical  flux qubit experiments  have values of $\alpha$ in the
range $0.6-0.9$ and $\eta$ in the range $0.1-0.6$.\cite{chiorescu,valenzuela}
In quantum computation implementations
\cite{qbit_mooij,chiorescu} the  FQ
is operated at  magnetic fields near the half-flux quantum,  $f=
1/2+\delta f$, with $\delta f \ll 1$. For $\alpha \ge
1/2$, the potential $V(\varphi_l,\varphi_t)$ has  
two minima 
at $(\varphi_l,\varphi_t)=(\pm\varphi^*,0)$ separated by a
maximum at $(\varphi_l,\varphi_t)=(0,0)$. Each minima corresponds
to macroscopic persistent currents of opposite sign, and 
for $\delta f\gtrsim 0$ ($\delta f \lesssim 0$) a ground state $|+\rangle$
($|-\rangle$) with positive (negative) loop current is
favored.

\begin{figure}[th] 
	\includegraphics[width=15pc]{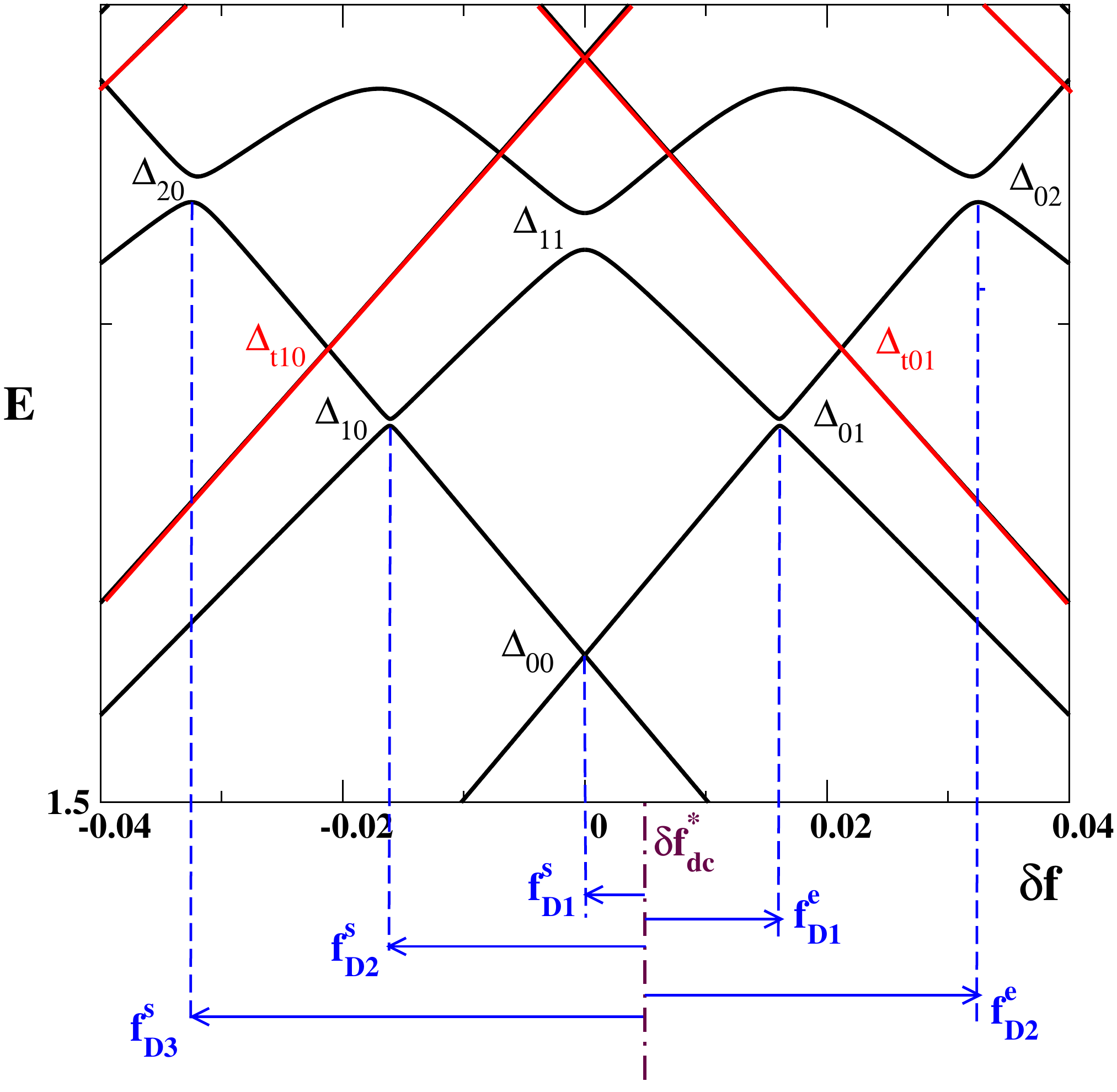}
	\caption{Lowest   energy levels  of the flux qubit as a function of
		flux detuning  $\delta f$, for the  qubit parameters $\eta=0.25$ and $\alpha=0.80$ considered throughout  this work. 
		Energy is measured in units of $E_J$ and flux in units
		of $\Phi_0$. The gaps at the avoided crossings are indicated as $\Delta_{ij}$. Black lines correspond to ``longitudinal'' modes and
		red lines to ``transverse'' modes (see text for description).
		The blue arrows at the bottom of the plot illustrate the relation between 
		ac driving amplitudes  and level crossing positions for a particular dc flux detunning $\delta f_{dc}^*$. The indicated values  correspond to the edges of the spectroscopic diamonds of Fig.\ref{las} for the given $\delta f_{dc}^*$.
		} \label{fig-espec}
\end{figure}

In Fig.\ref{fig-espec}  we plot the lowest 
energy levels $E_n$ as a function of $\delta f=f-1/2$, obtained  by numerical
diagonalization of  ${\cal H}_{FQ}$.\cite{foot1}  
In this case we set $\eta= 0.25$  and $\alpha=0.8$, close to the
experimental values  employed in flux qubits experiments.\cite{qbit_mooij,valenzuela} 
Negative (positive) slopes in Fig.\ref{fig-espec}, 
correspond to  eigenstates  with positive (negative) loop current.
A gap $\Delta_{ij}$ opens at the avoided crossings 
of the $i$-th level of positive slope with
the $j$-th level of negative slope at $\delta f=f_{ij}$. 
For our choice of device parameters, the avoided crossing
of  the two lowest levels at $\delta f=f_{00}=0$
has a  gap $\Delta_{00}=3.33\times 10^{-4}$
(in units of $E_J$).
At larger $\delta f$, we 
find  the avoided crossing with a 
third level at $f_{01}=-f_{10}=0.0161$ with gap $\Delta_{01}=\Delta_{10}=2.16\times 10^{-3}$, and
next the avoided crossing at  $f_{02}=-f_{20}=0.0323$ 
with gap  $\Delta_{02}=\Delta_{20}=8.26\times 10^{-3}$.
(There are also energy levels that correspond 
to excited transverse modes,\cite{foot2}
plotted with red lines in Fig.\ref{fig-espec},
but they  have a negligible contribution to
the dynamics considered here).

The two-level regime, involving only the lowest
eigenstates at $E_0(\delta f)$ and $E_1(\delta f)$,
corresponds to $|\delta f| \ll f_{01}$, such
that the avoided crossings with the third energy level are not reached. 
In this case,
the hamiltonian  of Eq.~(\ref{ham-sys}) can be reduced to a two-level system\cite{qbit_mooij,ferron} 

\begin{equation}\label{htls}
 {\cal H}_{TLS}=-\frac{\epsilon}{2} {\hat\sigma}_z - \frac{\Delta}{2} {\hat\sigma}_x
\;,
\end{equation}
in the basis defined by the persistent current states 
$|\pm\rangle=(|0\rangle\pm|1\rangle)/\sqrt{2}$, with $|0\rangle$ and $|1\rangle$
the ground and excited states at $\delta f=0$.
The parameters of ${\cal H}_{TLS}$ are the gap $\Delta=\Delta_{00}$
and the detuning energy
$\epsilon= 4\pi I_p \delta f$.
Here $I_p=\alpha |\langle+|\sin2\varphi _l|+\rangle|=
\alpha |\langle-|\sin2\varphi _l|-\rangle|$ is the magnitude of the loop current, 
which for our case with $\alpha=0.8$ and $\eta=0.25$
 is $I_p=0.721$ (in units of $I_c=2\pi E_J/\Phi_0$).

\subsection{Landau-Zener-St\"uckelberg interferometry}

Landau-Zener-St\"uckelberg (LZS) interferometry is performed by
applying an harmonic field on top of the static field  with
\begin{equation}\label{tdf}
\delta f\rightarrow  \delta f(t)=\delta f_{dc}+f_{ac}\sin{(\omega_{0} t)} \;.
\end{equation}
If the ac driving amplitude is such that $|\delta f_{dc} \pm f_{ac}|<f_{01}$, which is fulfilled
for $f_{ac}<f_{01}/2$, the quantum dynamics can be described within the two-level regime. 
In this case,   we use $\epsilon (t)= \epsilon_{0} + A \sin
(\omega_{0} t)$  in the Hamiltonian of  Eq.(\ref{htls}), 
with $\epsilon_{0}=4\pi I_p \delta f_{dc}$ and $A = 4\pi I_p  f_{ac}$.

When $f_{ac}>|\delta f_{dc}|$ the central avoided crossing at $\delta f=0$ 
is reached within the range of the driving amplitude, $\delta f_{dc} \pm f_{ac}$. In this case
the periodically repeated Landau-Zener transitions at $\delta f=0$ give place to LZS 
interference patterns as a function of $\delta f_{dc}$  and $f_{ac}$, which are characterized
by {\it multiphoton resonances}\cite{shirley} and {\it coherent destruction of tunneling},\cite{grossmann} as we describe below.

\begin{figure}[th]
	\begin{center}
		\includegraphics[width=0.9\linewidth,clip]{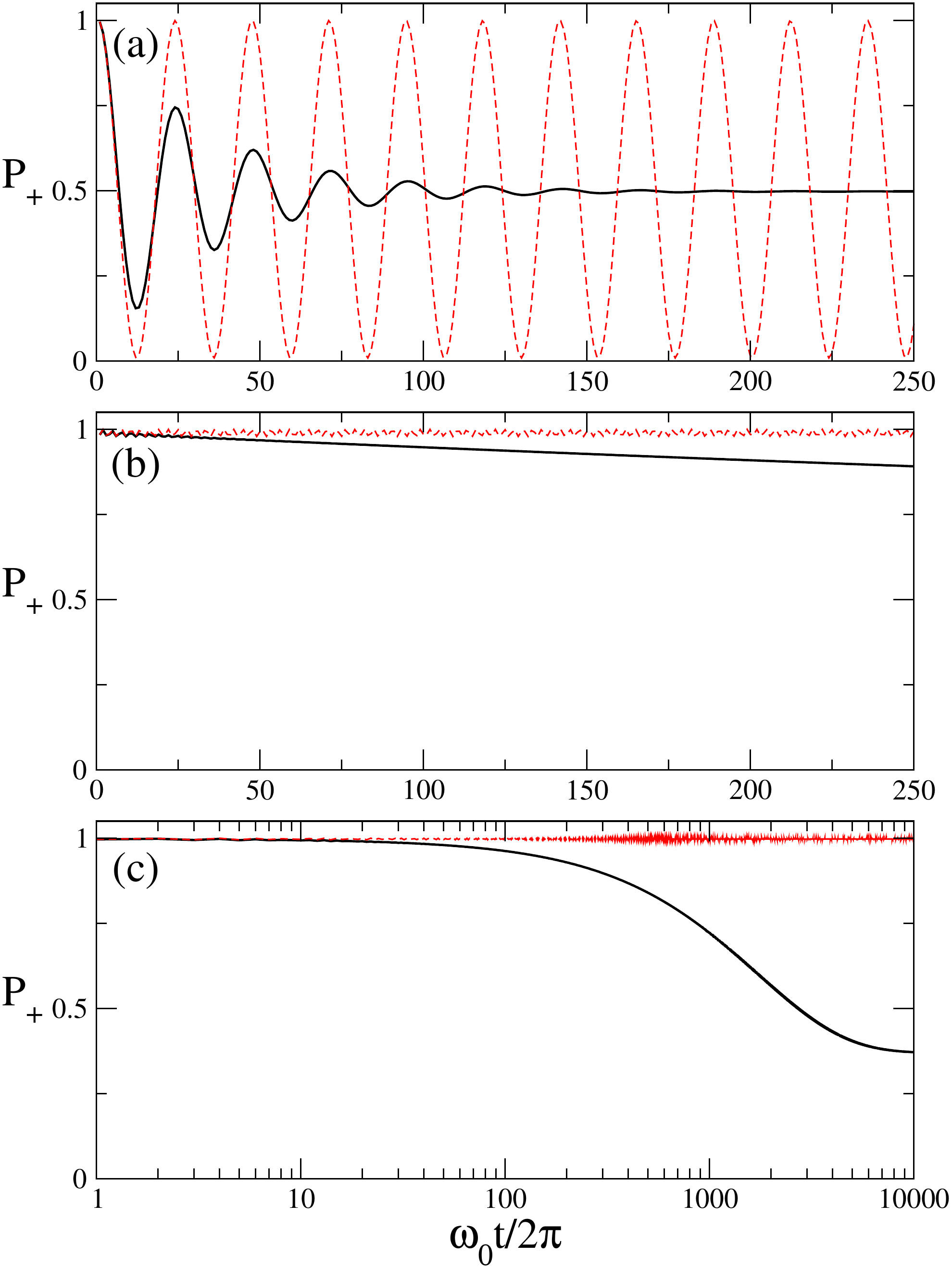}
	\end{center}
	\caption{(Color online)
		 Probability of a positive loop current $P_+$ state as a function of time,
		 for a flux qubit driven with frequency $\omega_0=0.003$ 
		 (in units of $E_J/\hbar$).
		 (a) $\delta f_{dc}=0.00066$ (second resonance)
		and $f_{ac}=0.0007$. (b) $\delta f_{dc}=0.00151$ (off resonance)
		and $f_{ac}=0.00245$. (c) Same as (b) but for a long time scale.  
		Red dashed line shows the results in the non dissipative
		case and the black line corresponds to the dissipative case with an ohmic bath 
		at $T=20\;mK$.
	} \label{fo1}
\end{figure}

There are {\it multiphoton resonances} when\cite{shirley}
$E_1(\delta f)-E_0(\delta f)=n\omega_{0},$
(where $\omega_0$ is written in units of $E_J/\hbar$ and energies in 
units of $E_J$).
If $\omega_0\gg \Delta$, these resonances are at $\epsilon_0= n\omega_0$.\cite{shevchenko,grifoni-hanggi,grifonih,ashhab} 
Calling $f_\omega = \omega_0/4\pi I_p$, 
the $n$-resonance condition can also be written as $\delta f _{dc}=nf_\omega$.
An example of the dynamic behavior in a $n=2$ resonance is shown in Fig.\ref{fo1}(a),
calculated as described in Ref.~\onlinecite{fds} (see also the Appendix for details of the calculation).
In this case the FQ is driven with frequency $\omega_0=0.003$ (for
$E_J/h\sim 300$GHz it corresponds to $\omega_0/2\pi\sim 900$Mhz).
The dc detuning  is $\delta f_{dc}  = 0.00066$, corresponding to $\delta f_{dc}/f_\omega = 2$
(since $f_\omega=\omega_0/4\pi I_p=0.00033$).
The FQ is started at $t=0$ in the ground state $|\Psi(t=0)\rangle=|0\rangle$ ,
and the probability of having a positive loop current is calculated, $P_+(t)=|\langle\Psi(t)|+\rangle|^2$.
Since for $\delta f_{dc}>0$ we have $|0\rangle\approx|+\rangle$,
the initial probability is $P_+(t=0)\approx 1$. 
The red-dashed line in Fig \ref{fo1}(a) shows the time dependence of $P_+(t)$.
The resonant dynamics is clearly seen: the FQ oscillates coherently between positive and negative
current states. Therefore, the probability $P_+(t)$  oscillates between $0$ and $1$, 
having a time averaged value of $\overline{ P_+}=1/2$.
In contrast, Fig.\ref{fo1}(b) shows the time dependence of $P_+(t)$ in an
off-resonant case, for $\delta f_{dc}/f_{\omega}=4.58$.
We see that for off-resonance the FQ stays in the positive loop current
state, since $P_+(t)$ fluctuates around  $P_+(t)\lesssim 1$.

For $A\omega_{0} \gg \Delta^{2}$, it has been shown, in a rotating wave approximation, that 
the average probability  $\overline{ P_+}$  near a $n-$ photon resonance is\cite{shevchenko,ashhab,grifonih,oliver}
\begin{equation}
\overline{P_{+}}=1-\frac{1}{2} \frac{\Delta_{n}^{2}}
{(n \omega_{0} - {\epsilon}_{0})^{2} + \Delta_{n}^{2}}  \; . \label{l2ls}
\end{equation}
 At the resonance, $\epsilon_0=n\omega_0$, Eq.(\ref{l2ls})
gives $\overline{P_{+}}=1/2$ and away of the resonance $\overline{P_{+}}\lesssim 1$.
The width of the resonance is  $\delta\epsilon=|\Delta_n|=\Delta |J_{n} (A/ \omega_{0})|=\Delta |J_{n} (f_{ac}/f_\omega)|$, with  $J_n(x)$ the Bessel function of the first kind. 
This gives a quasiperiodic dependence as a function of $f_{ac}$ for $\delta f_{dc}$ fixed
near the resonance. In particular, at the zeros of $J_n(x)$,
the resonance is destroyed, giving $\overline{ P_+}=1$ instead of $\overline{ P_+}=1/2$, a 
phenomenon known as {\it coherent destruction of tunneling}.\cite{grossmann,kayanuma} Plots of $\overline{P_{+}}$
as a function of flux detuning $\delta f_{dc}$ and ac amplitude $f_{ac}$ give the typical 
 LZS interference patterns, which have been measured experimentally by Oliver et al
in flux qubits,\cite{oliver,berns1,rudner,valenzuela} and have also been observed in other systems.
\cite{izmalkov,cqubits,wilson,wilson2,sun,sun2,wang2010,graaf2013,shevchenko2,mark,petta2010,stehlik2012,cao2013,dupont2013,shang2013,nalbach2013,ludwig2014,stace2014,granger2015,huang,zhou2014,lahaye,kling}

The Eq.(\ref{l2ls}) corresponds to ideally isolated flux qubits, neglecting the coupling
of the qubit  with the environment. The dynamics of the FQ  as an open quantum system is 
usually characterized by the energy relaxation time $t_r \equiv T_1 $ and the decoherence time $t_{dec}\equiv T_2 $.
Several phenomenological approaches have taken into account relaxation and decoherence 
in LZS interferometry, obtaining a broadening of the Lorentzian-shape $n-$ photon resonances of Eq.(\ref{l2ls}). 
For example, in Ref.\onlinecite{shevchenko} a Bloch equation approach is used, obtaining for
zero temperature:
\begin{equation}
	\overline{P_{+}}=1-\frac{1}{2}
	\frac{\frac{\Gamma_2}{\Gamma_1}\Delta_{n}^{2}}
	{(n \omega_{0} - {\epsilon}_{0})^{2} + \frac{\Gamma_2}{\Gamma_1}\Delta_{n}^{2}+{\Gamma_2}^2}  \; . \label{l2lsr}
\end{equation}
with $\Gamma_1=1/T_1$ and $\Gamma_2=1/T_2$.
Similar results were obtained by  Berns et al,\cite{berns1}
 considering a Pauli rate equation 
with  an effective transition rate that adds to $\Gamma_1$ a driving induced
rate $W$. The latter was obtained assuming that decoherence is due to a classical white noise in the magnetic flux,
an approach that is valid for time scales smaller than the relaxation time and larger than the decoherence time,  for $T_1>t\gg T_2$.
The case of low frequency $1/\omega$ noise has been considered in a similar approach,\cite{yu2010a,yu2010b}
finding a Gaussian line shape for the $n$-resonances
instead of the Lorentzian line shape.

\section{Dynamic transition in LZS interference patterns}

\subsection{The driven flux qubit with an Ohmic bath}

The description of the LZS interference patterns with phenomenological
approaches like Eq.(\ref{l2lsr}) gives a good agreement with most
of the experimental results.\cite{valenzuela,wen} However, they rely on approximations
valid either for large frequencies, or for low ac amplitudes, or
for time scales smaller than the relaxation time.
Here we will study LZS interferometry using the Floquet formalism
for time-periodic Hamiltonians,\cite{shirley,grifoni-hanggi,grifonih,fds} which allows for an 
exact treatment of driving forces  of arbitrary strength and frequency.
The time-dependent  Schr\"odinger equation 
is transformed to an equivalent  eigenvalue equation for the Floquet states
$|\alpha(t)\rangle$ and  quasi-energies $\varepsilon_\alpha$,
which can be solved numerically. (See the Appendix for details).
To describe relaxation and  decoherence processes, 
we consider that the qubit is  weakly coupled to  a bath of harmonic oscillators.
The Markov approximation of the bath correlations is performed after
writing the reduced density matrix $\hat\rho$ of the qubit 
in the Floquet basis,
$\rho_{\alpha\beta}(t)=\langle\alpha(t)|\hat\rho(t)|\beta(t)\rangle$.\cite{grifoni-hanggi,grifonih,fds2}
By following this procedure, the quantum master equation obtained
for $\rho_{\alpha\beta}$
is valid for periodic driving forces   of arbitrary strength.
(In contrast, the standard Born-Markov approach is valid for
small driving forces). 
The obtained Floquet-Markov quantum master equation is  
\cite{grifoni-hanggi,grifonih,fds2}:
\begin{eqnarray}
\label{drho}
\frac{d\rho_{\alpha\beta}(t)}{dt}&=&
\sum_{\alpha'\beta'} \Lambda_{\alpha\beta\alpha'\beta'}\;\rho_{\alpha'\beta'}\,,
\label{FM0}\\
\Lambda_{\alpha\beta\alpha'\beta'}&=&-\frac{i}{\hbar}
(\varepsilon_\alpha-\varepsilon_\beta)\delta_{\alpha\alpha'}\delta_{\beta\beta'}+ L_{\alpha\beta\alpha'\beta'}\;.
\end{eqnarray}
The coefficients  ${\it L}_{\alpha\beta\alpha'\beta'}$
depend on the spectral density $J(\omega)$ of the bath,  the temperature $T$, and
on the qubit-bath coupling.
Using the numerically obtained Floquet states $|\alpha(t)\rangle$,
we can calculate the coefficients  ${\it L}_{\alpha\beta\alpha'\beta'}$ ,
and then we compute the solution of $\rho_{\alpha\beta}(t)$ as
described in the Appendix.

Here, we will consider the dynamics of a FQ coupled 
to a bath with an ohmic spectral  density\cite{grifoni-hanggi,grifonih,fds2} 
$$J(\omega)=\gamma\omega e^{-\omega/\omega_c} ,$$
with $\omega_c$ a cutoff frequency.
The Ohmic bath mimics an unstructured 
electromagnetic environment that in the classical limit leads to white noise.
We consider $\gamma=0.001$, 
corresponding to weak dissipation,\cite{oliver,valenzuela}
and a large cutoff
$\omega_c = 1.0 E_J/\hbar \gg \omega_{0}$.
The bath temperature is taken
as  $T= 0.0014 E_J/k_B$ ($\sim 20{\rm mK}$ for $E_J/h 
\sim 300{\rm GHz}$).

Experimentally, the probability of having a state of positive or negative
persistent current in the flux qubit is measured.\cite{chiorescu,oliver,valenzuela} 
The probability of a positive loop current measurement 
can be calculated in general as\cite{fds,fds2} 
$$P_+(t)={\rm Tr}[\hat\Pi_+ \hat \rho(t)]$$ 
with $\hat\Pi_+$ the operator that projects
wave functions on the $\varphi_l>0$ subspace,
as described in the Appendix. 
In Fig \ref{fo1} the black lines show
the population $P_+$ as a function of time obtained
from the numerical solution of Eq.(\ref{drho}), taking the ground state 
$|0\rangle$ as initial condition, and compared with the isolated
FQ (red dashed lines) discussed in the previous section. 
In Fig \ref{fo1}(a), for the  $n = 2$ resonance, we see that
the population $P_+(t)$ has damped oscillations  that
tend to the asymptotic average value of $\overline{ P_+}=1/2$ for large times.
On the other hand, for the off-resonant case of Fig \ref{fo1}(b) and (c) 
we see that there is a clear difference between the short time
behavior of $P_+(t)$, shown in Fig \ref{fo1}(b), and
the large time behavior shown in Fig  \ref{fo1}(c).
Moreover,  we observe that 
$P_+(t)$  tends to asymptotic values that are  very different
than in the isolated system.
For the particular case shown in the plot,
we find population inversion, i.e $P_+ < 1/2$,
in the asymptotic long time limit, a phenomenon
we discussed in Ref.\onlinecite{fds2} for off-resonant cases.
In the following, we will see how this change of behavior in
the long time is reflected in  the full LZS interference pattern
as a function of $\delta f_{dc}$ and$f_{ac}$.

\begin{figure}[h t]
	\begin{center}
		\includegraphics[width=\linewidth]{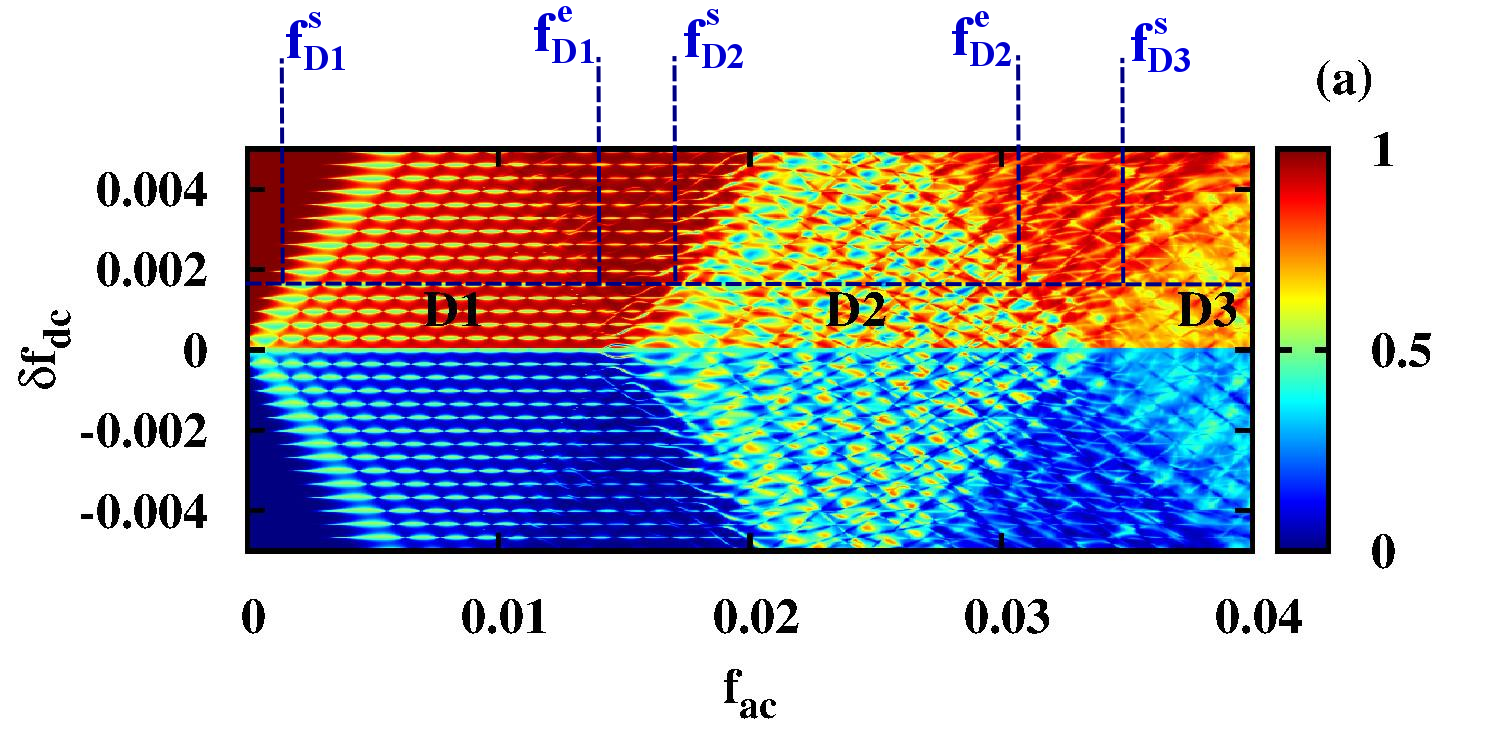}		
       \includegraphics[width=\linewidth]{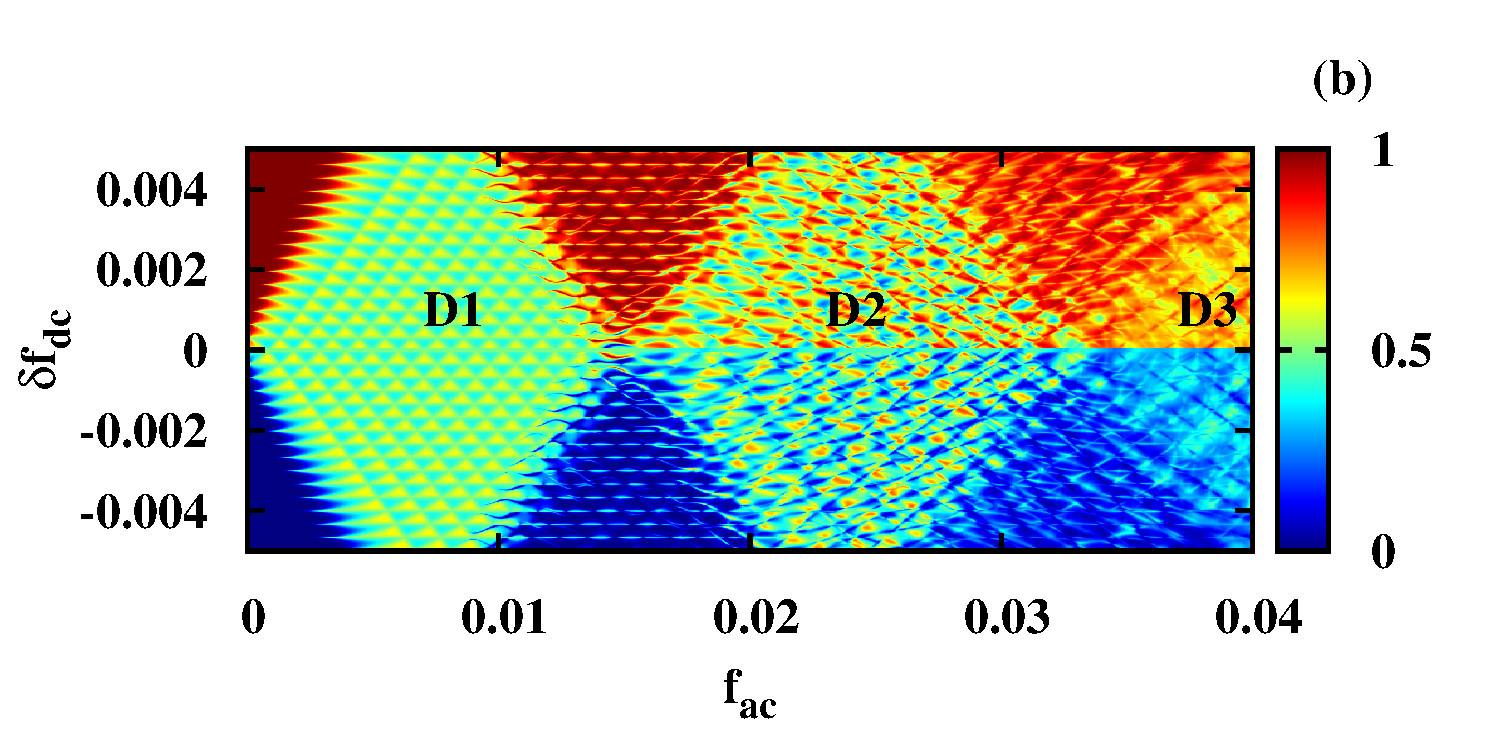}			
		\includegraphics[width=0.8\linewidth,clip]{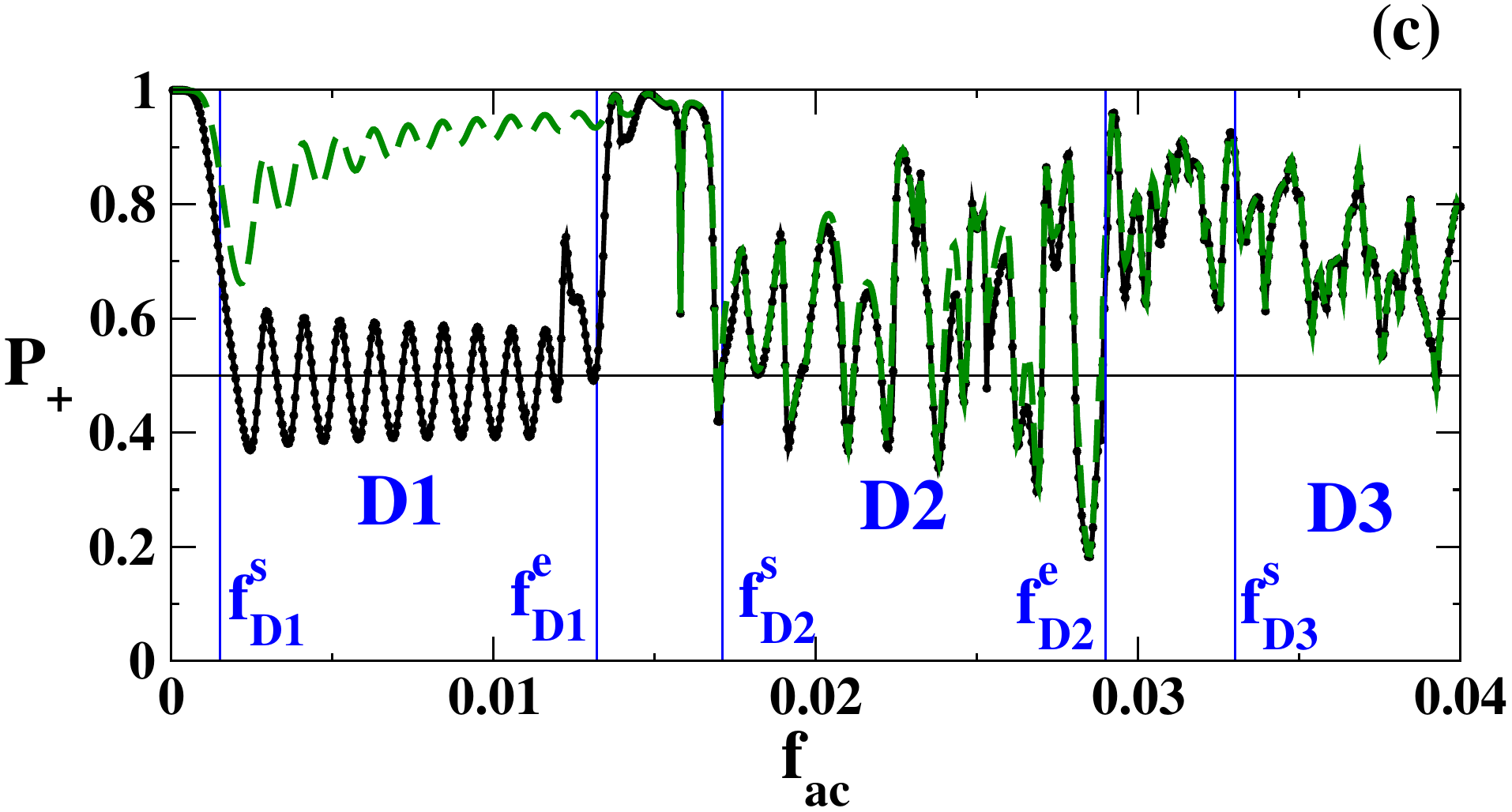}
\end{center}
	\caption{Landau-Zener-Stuckelberg interference patterns. Showed are
		intensity plots of $P_+$ as a function of the driving amplitude $f_{ac}$
		and dc detuning $\delta f_{dc}$ for (a) $t=1000 \tau$, time scale
		of the experiments,  and (b) asymptotic regime for $t\rightarrow\infty$.
		In (a) we mark the ac amplitudes defined in the text that correspond to the  edges of the spectroscopic diamonds for a given $\delta f_{dc}^*$.
		(c) $P_+$ for $\delta f_{dc}=0.00151$ as a function of the driving amplitude $f_{ac}$, at $t=1000 \tau$ (blue dashed line) and asymptotic 
		average population $\overline P_+$ (black line) .
		The calculations were performed for
		$\omega_{0}= 2\pi/\tau= 0.003 E_J/\hbar$ and
		ohmic bath  at $T=0.0014 E_J/k_B\sim 20$mK.
		Vertical lines separate diamond regimes D1, D2, D3, described in the text. 
	} \label{las}
\end{figure}

\subsection{Dynamic transition in LZS interferometry}

In Figs.\ref{las}(a) and \ref{las}(b) we show an intensity plot of 
$P_+$ as a function of $\delta f_{dc}$ and $f_{ac}$,  calculated
at a time 
near the experimental time scale of Refs.\onlinecite{oliver,berns1,valenzuela},
$t_{exp}=1000\tau$, with $\tau=2\pi/\omega$ the period of the ac driving. 
For $f_{ac}<f_{01}/2$ we observe an LZS pattern modulated by
multiphoton resonances and coherence destruction of tunneling,
which can be described within a good approximation with
Eq.(\ref{l2lsr}).
This LZS interference plot  is very similar
to the experimental results of Ref.\onlinecite{oliver,valenzuela}.
Furthermore, for higher ac amplitudes,  we find
a pattern of  ``spectroscopic diamonds'', also
in good agreement with experiments.\cite{valenzuela} 
The  diamond structure obtained for high ac amplitudes
can be related to the energy level spectrum of Fig.\ref{fig-espec},
for a fixed dc flux detuning $f_{01}>\delta f_{dc}>0$,
as follows.\cite{valenzuela,wen}
The first diamond, D1, 
starts when the $\Delta_{00}=\Delta$ avoided crossing at $\delta f = f_{00}=0$ 
is reached by the amplitude of the ac drive, {\it i.e.} when
$f_{00}=\delta f_{dc}\pm f_{ac}=0$.
This defines  an onset ac amplitude $f_{D1}^{s}=\delta f_{dc}$. 
The first diamond ends when the nearest second avoided crossing 
is reached (with gap $\Delta_{01}$) 
at the ac amplitude $f_{D1}^{e}=f_{01}-\delta f_{dc}$. 
Then the second diamond, D2, starts when the other second 
avoided crossing is reached, at 
$f_{D2}^{s}=\delta f_{dc}-f_{10}$, 
and it ends when the next avoided crossing is reached at
$f_{D2}^{e}=f_{02}-\delta f_{dc}$.
Similarly, the third diamond, D3, starts at 
$f_{D3}^{s}=\delta f_{dc}-f_{20}$,
and so on.
The analysis of the positions of the resonances  as a function
of $f_{ac}$ and  $\delta f_{dc}$ was the route followed 
in Refs.\onlinecite{valenzuela}
 to obtain the parameters characterizing the different avoided crossings
of the flux qubit.

The FQ of Refs.\onlinecite{oliver,valenzuela} have short decoherence times
($t_{dec}\sim 20{\rm ns} > \tau = \frac{2\pi}{\omega_0} \sim 1-10{\rm ns}$)
and large relaxation times ($t_r\sim 100\mu{\rm s}$). 
The typical duration  of the driving pulses in these experimental measurements 
($t_{exp}\sim 3 \mu{\rm s}$) is in 
between this two time scales, $t_{dec} < t_{exp} < t_r $.
Due to this time scale separation,
a model with classical noise,\cite{berns1} valid for $t\ll t_r$, 
can  qualitatively explain the experimentally observed behavior of $P_+$
within the first diamond, through  Eq.(\ref{l2lsr}).\cite{oliver,shevchenko}
Moreover, a multilevel extension of the model of Ref.\onlinecite{berns1}
can also describe the higher order diamonds (provided one gives as an input parameter
the positions $f_{ij}$ and the gaps $\Delta_{ij}$ of the avoided crossings).\cite{wen}

For large times $t\gg t_r$ one would expect naively, that
after full relaxation with the environment, 
a blurred picture of the LZS interference pattern of 
Figs.\ref{las}(a) 
with broadened resonance lobes should be observed.
The asymptotic ${\overline P_+}\equiv 
{\lim}_{t\rightarrow\infty}\langle P_+(t)\rangle_\tau$, averaged
over one period $\tau$, is shown in Fig.\ref{las}(b). 
[
${\overline P_+}$ can be calculated exactly 
after obtaining numerically
the right eigenvector of  $\Lambda_{\alpha\beta\alpha'\beta'}$  with zero eigenvalue.
See the Appendix for details.]
Surprisingly, we see in  Fig.\ref{las}(b) that the asymptotic 
behavior of $P_+$ gives a qualitatively different LZS interference pattern
within the first diamond,
and not a mere blurred version of Fig.\ref{las}(a).
In particular, a cut at constant $\delta f_{dc}$ is shown
in Fig.\ref{las}(c). There we see clearly that
within the first diamond regime, $f_{D1}^s<f_{ac}<f_{D1}^e$, the value of $P_+$ at
the experimental time scale is very different than the asymptotic value 
${\overline P_+}$.  On the other hand, 
beyond the first diamond, for $f_{ac}>f_{D1}^e$ the
results of $P_+(t_{exp})$ and ${\overline P_+}$ are nearly coincident.

\begin{figure}[th]
	\begin{center}		
	\includegraphics[width=0.9\linewidth,clip]{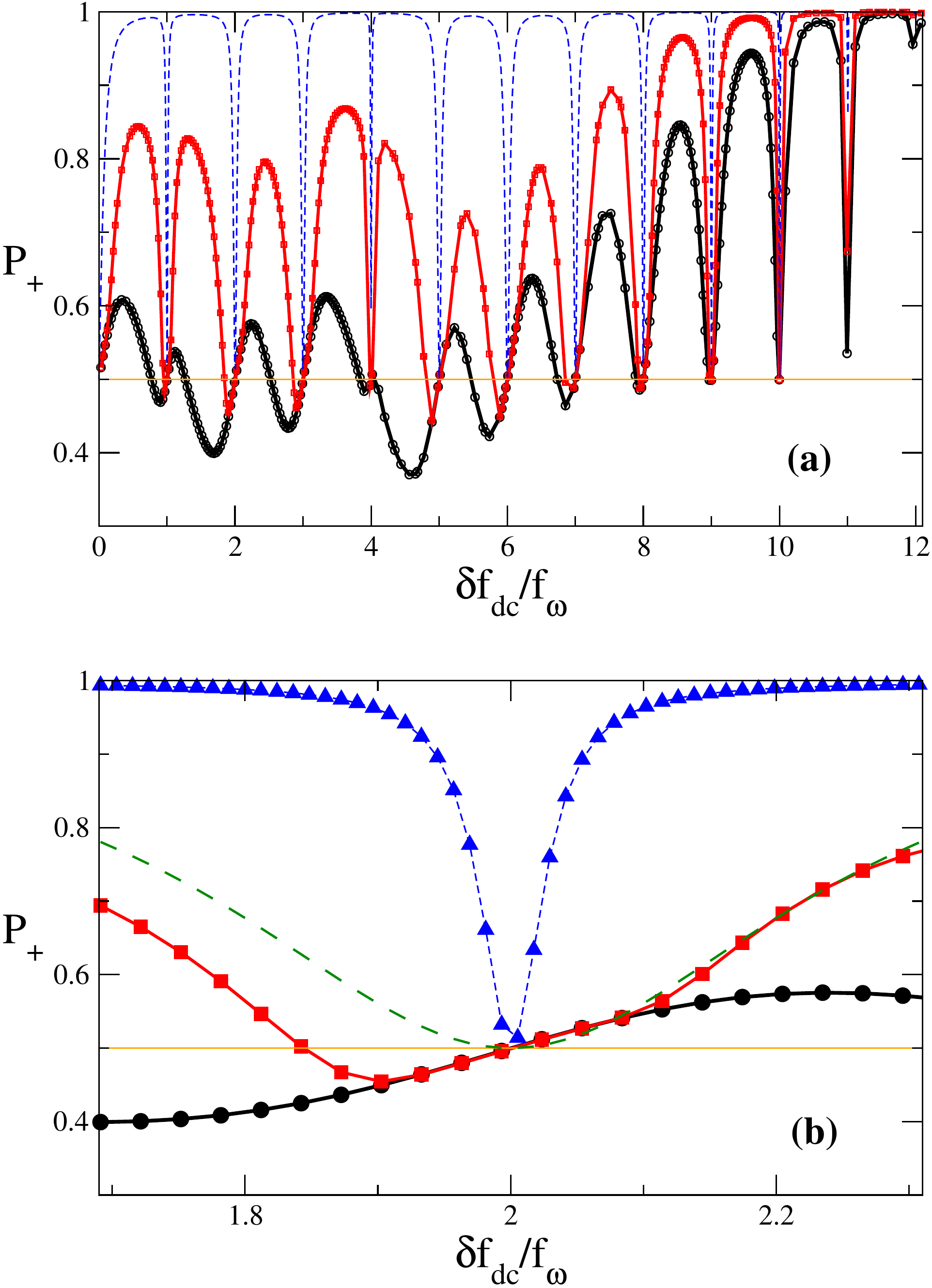}
	\end{center}
	\caption{(color online)
		(a) Population $P_+$ as a function of the $dc$ flux detuning $\delta f_{dc}$
		for the flux qubit driven with
		$f_{ac}=0.00245$ and $\omega_{0}= 2\pi/\tau= 0.003 E_J/\hbar$,
		and coupled to an ohmic bath 
		at $T=20$mK (for $E_J/h\approx300{\rm GHz}$).
		Red line with squares: $t=1000 \tau$. Black
		line with circles:  asymptotic ($t\rightarrow\infty$)
		average population $\overline P_+$.
		Blue line with triangles: time averaged population in the isolated system. Horizontal orange line: indicates  the 
		$P_+=0.5$ level to help to
		identify when there is population inversion.
		The flux detuning $\delta f$ is normalized by $f_\omega=\omega_0/4\pi I_p$, such that the $n$-photon resonances are at $\delta f=nf_\omega$.
		(b) Enlarged view around the $n=2$ resonance. The green dashed line is a plot
		of the best fit with Eq.(\ref{l2lsr}), while a plot of Eq.(\ref{l2ls}) (not shown) falls exactly over the blue dotted line.  
	} \label{fPf0}
\end{figure}

In Fig.\ref{fPf0} we show $P_+$ vs. $\delta f_{dc}$, for
$f_{ac}=0.00245$, within the first diamond.
The blue line with triangles corresponds to the solution of the isolated system,
which shows dips where $P_+\approx 1/2$ that  
correspond to the $n$-photon resonances at $\delta f_{dc}/f_\omega=n$.
The red line with squares 
corresponds to $P_+(t_{exp})$, which shows 
values of $P_+$ smaller than in the isolated case (due to effect of decoherence and relaxation) and in agreement with  the experiments. The asymptotic 
${\overline P_+}$ is also shown in Fig.\ref{fPf0} (black line with circles), 
which is  lower than $P_+(t_{exp})$.  Fig.\ref{fPf0}(b) shows
in detail the behavior near the $n=2$ resonance. The isolated
case shows a dip with a Lorentzian shape accurately described by
Eq.(\ref{l2ls}).  The open system  at the experimental time scale
shows a broadened peak for $P_+(t_{exp})$, partially consistent 
with a description like Eq.(\ref{l2lsr}), except that  near the
resonance the behavior of $P_+$ vs $\delta f_{dc}$   
becomes antisymmetric around $\delta f_{dc}/f_\omega=n$. In the asymptotic
steady regime, ${\overline P_+}$ vs $\delta f_{dc}$ is  antisymmetric in a wide
region around $\delta f_{dc}/f_\omega=n$, 
showing  population inversion (${\overline P_+}<1/2$)
below the resonance, for $\delta f_{dc}/f_\omega\lesssim n$.

\begin{figure}
	\begin{center}	
		\subfigure[\;$T=1.4$ mK]{\includegraphics[width=0.45\linewidth,clip]{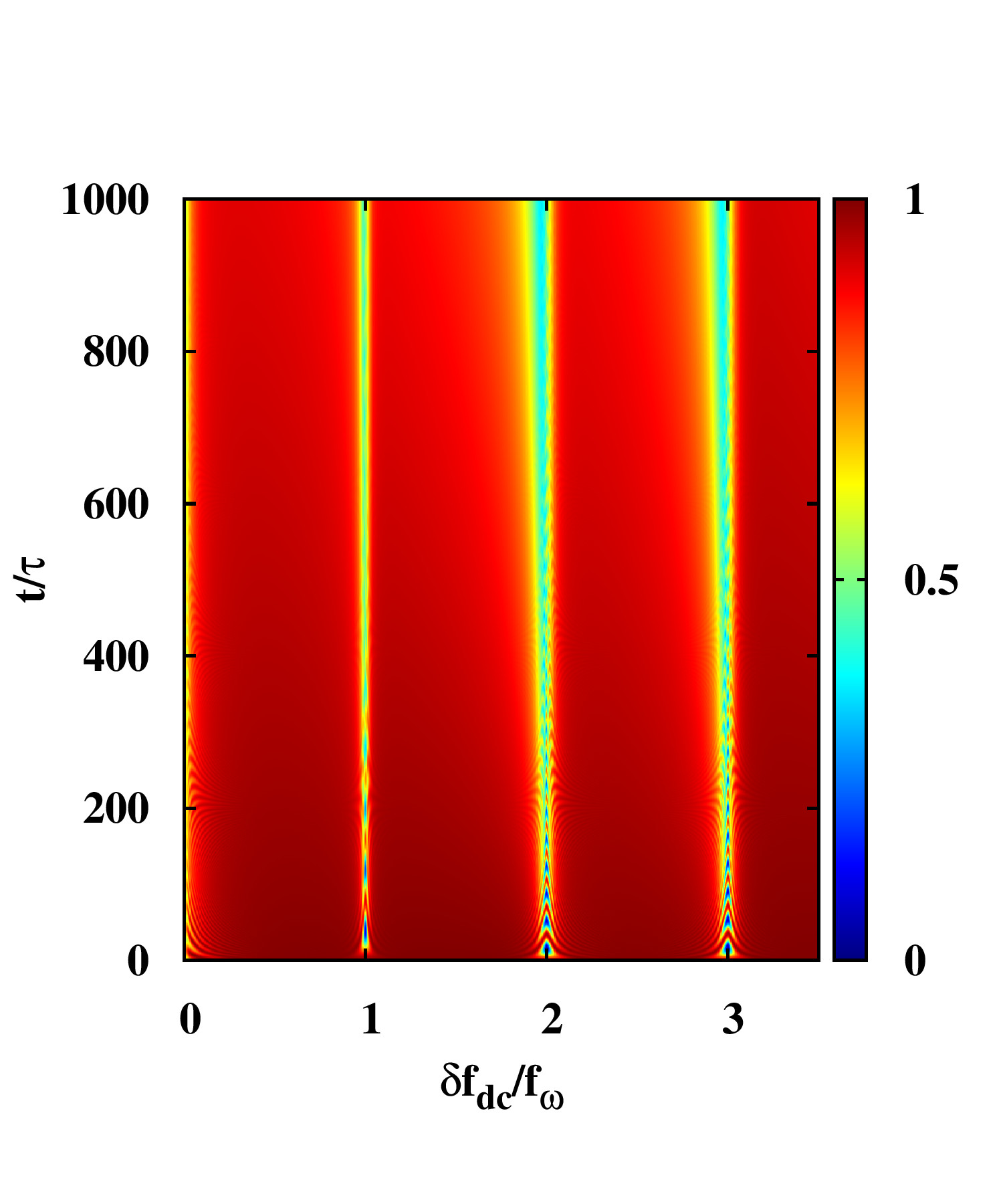}}
		\subfigure[\;$T=20$ mK]{\includegraphics[width=0.45\linewidth,clip]{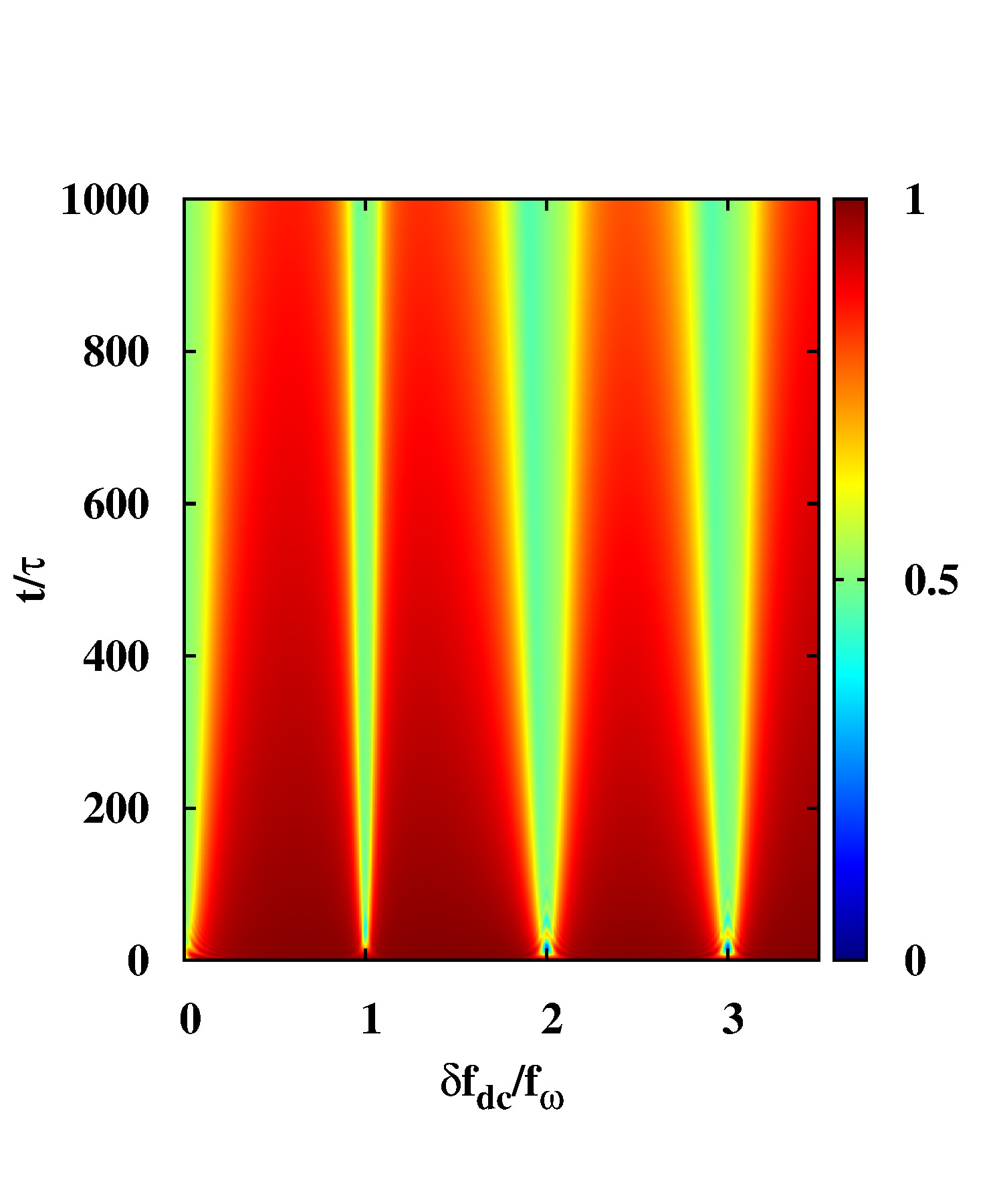}}\\
		\subfigure[\;$T=1.4$ mK]{\includegraphics[width=0.45\linewidth,clip]{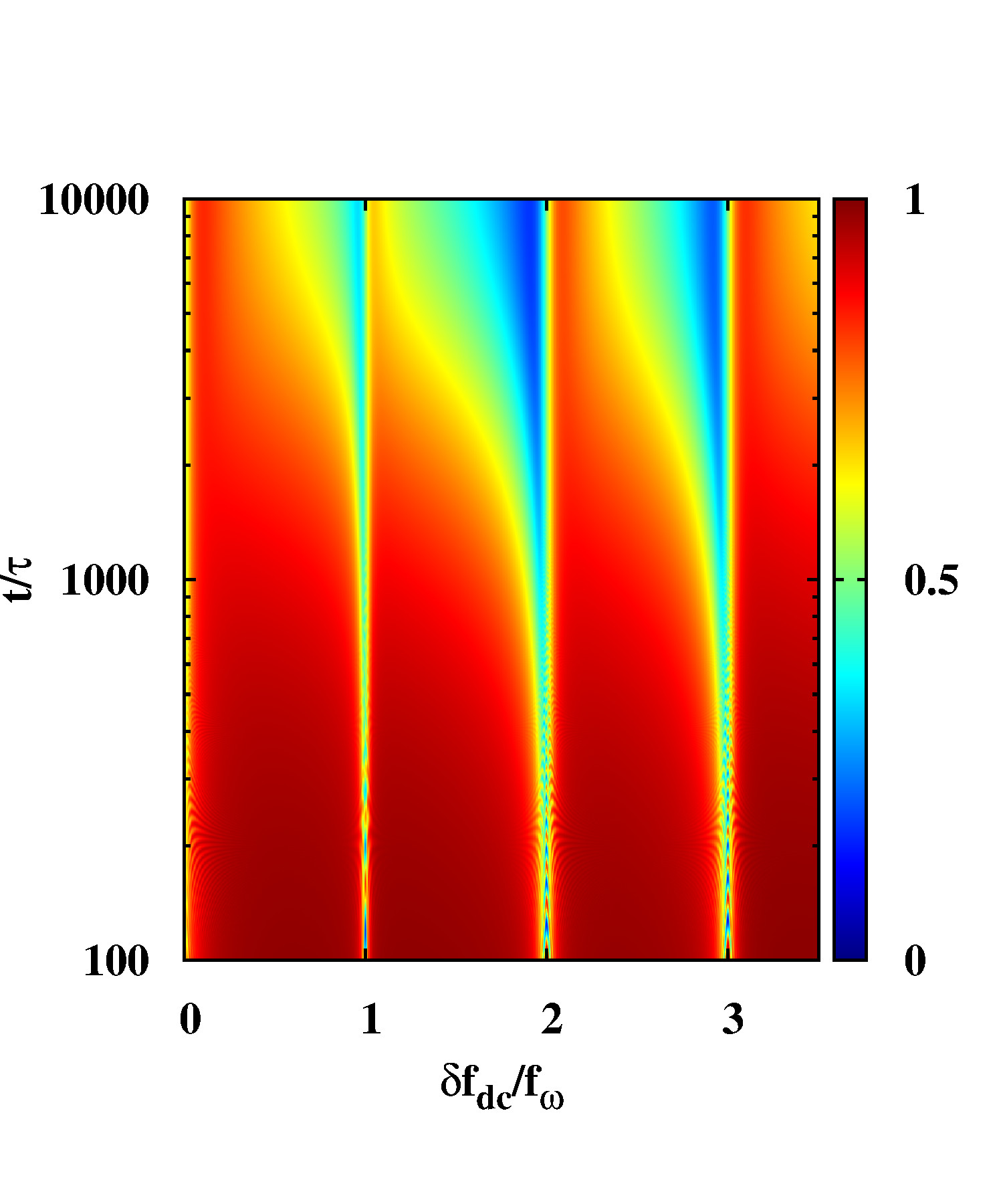}}
		\subfigure[\;$T=20$ mK]{\includegraphics[width=0.45\linewidth,clip]{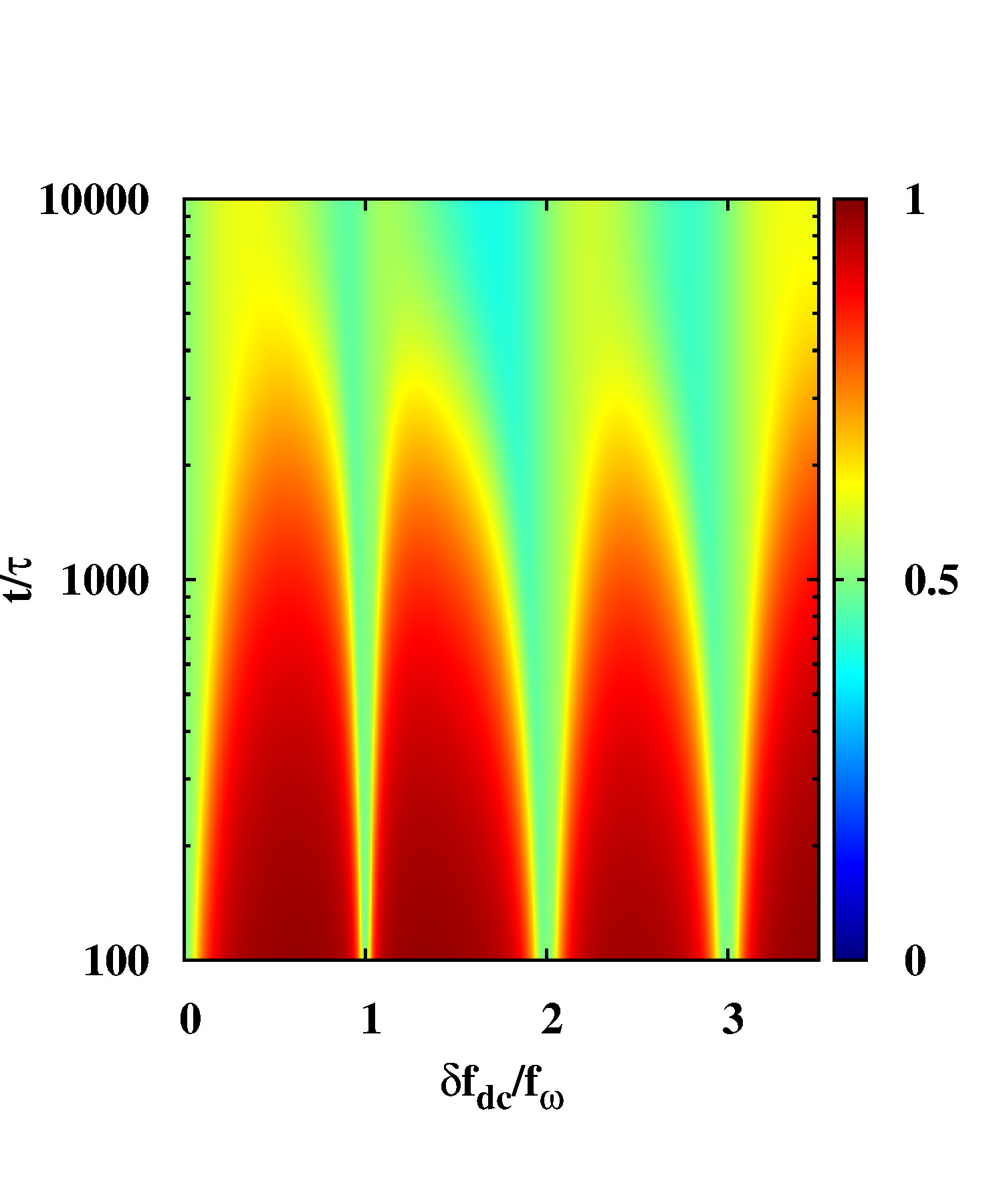}}	
	\end{center}
	\caption{(color online)  Intensity plots of the population $P_+$ 
as a function of  $\delta f_{dc}$ and driving time $t$
for the flux qubit driven with $f_{ac}=0.00245$ and $\omega_{0}= 0.003 E_J/\hbar$, and different temperatures. 
In (a) and (b) time is plotted in linear scale up to  times of the
order of the duration of the driving pulses in the 
experiments. In (c) and (d) time is plotted in logarithmic scale
showing the dynamic transition to antisymmetric resonances for large times,
when $t\gg t_r$. The flux detuning $\delta f_{dc}$ is normalized by $f_\omega=\omega_0/4\pi I_p$, such that the $n$-photon resonances are at $\delta f=nf_\omega$, and time is normalized by the driving period $\tau=2\pi/\omega_0$.
$T$ values are given in mK, 
corresponding to devices with $E_J/h\approx300{\rm GHz}$.}
\label{figtemp}
\end{figure}

The temporal evolution from symmetric to antisymmetric resonances can
be seen in detail in Fig.\ref{figtemp}, where we
show an intensity plot of 
$P_+$ as a function of $\delta f_{dc}$ and the duration time $t$, 
for $f_{ac}=0.00245$. 
Two different temperatures are considered,
a low temperature $T=1.4mK < \Delta$ in Figs.\ref{figtemp}(a) and (c), and
$T = 20mK > \Delta$ in Figs.\ref{figtemp}(b) and (d).
 The relaxation time $t_r$ for the driving amplitude considered in   Fig.\ref{figtemp} is $t_r\approx 1720\tau$
 (as we will see in Sec.IIIC, $t_r$ depends on $f_{ac}$).   
Figs.\ref{figtemp}(a) and (b) are for duration times up to the
time scale  $t_{exp} \sim 1000\tau$ of the experiments, for the two different temperatures. We
see that within this time scale $t_{exp}<t_r$, the resonances are
symmetric. 
In Figs.\ref{figtemp}(c) and (d) the duration time is plotted in
logarithmic scale and the evolution at very large times is shown.
We see  that at a time $t\sim t_r$ the resonances start to change form, 
becoming asymmetric in the long time limit. The asymmetry is stronger
for low temperatures, in particular for $T < \Delta$
there is full inversion of population 
on one side of the resonances.

\begin{figure}
	\begin{center}
		\includegraphics[width=0.98\linewidth,clip]{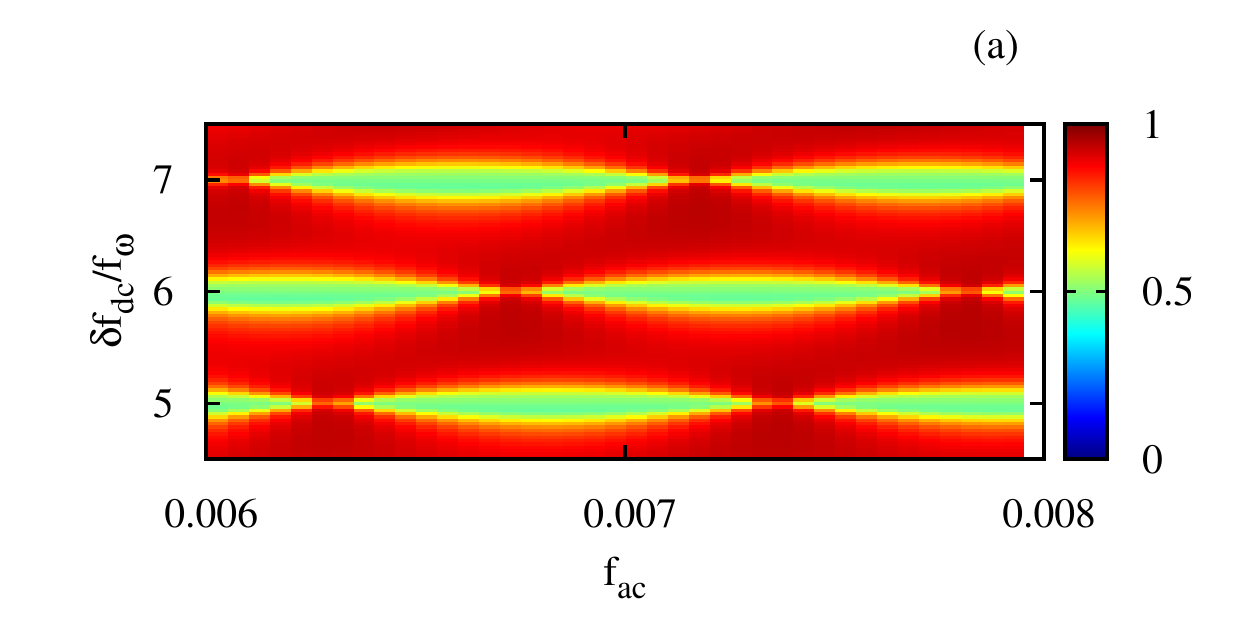}
		\includegraphics[width=0.98\linewidth,clip]{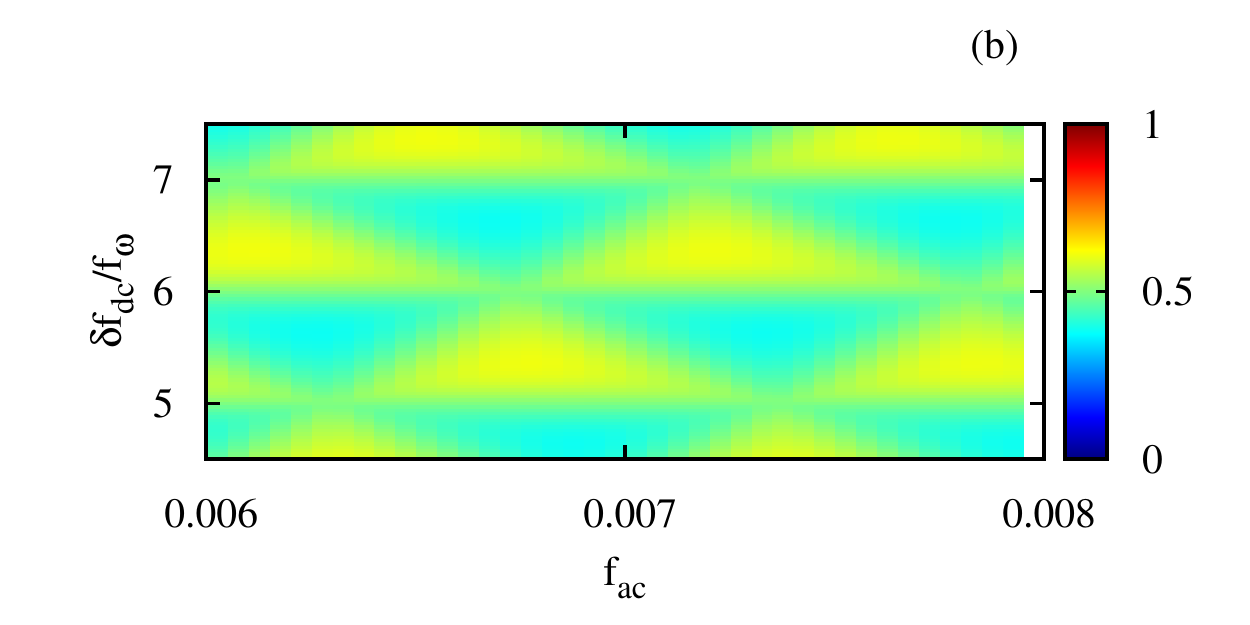}
	\end{center}
	\caption{(Color online)
		Dynamic transition in the LZS interference pattern from: (a) resonance lobes for $t=1000\tau\sim t_{exp}$, to (b) triangular checkerboard pattern for  asymptotic 
		long times. Enlarged views of Fig.\ref{las}(a) and (b), respectively, showing a sector of the first diamond including the  $n=5,6,7$ multiphoton resonances.
		The flux detuning $\delta f$ is normalized by $f_\omega=\omega_0/4\pi I_p$, such that the $n$-photon resonances are at $\delta f=nf_\omega$.
	} \label{las2}
\end{figure}

Fig.\ref{figtemp} shows that, as a function of time,
{\it there is a dynamic transition manifested by a symmetry change in the structure
of the LZS interference pattern}. 
The dynamic transition is from nearly symmetric resonances at short times ($t\ll t_r$) to {\it antisymmetric resonances} at long times ($t\gg t_r$).  
This is clearly illustrated in Fig.\ref{las2}
where  an enlarged view of a part of the first diamond is shown.
Fig.\ref{las2}(a) corresponds to $P_+(t_{exp})$ and shows the characteristic features
observed in experiments: (i) $P_+\approx 1$ away from the resonances,
(ii)  there are resonance lobes where $P_+\approx0.5$, and 
(iii) the $n$-resonance lobes are limited by the points where there is coherent destruction
of tunneling (CDT),  given by  $J_n(f_{ac}/f_\omega)=0$.
On the other hand,
in the asymptotic regime, [(Fig.\ref{las2}(b)], 
the pattern of symmetric $n$-resonance lobes is replaced by a pattern of
antisymmetric resonances, 
which form a triangular checkerboard
picture defined by triangles with $P_+<0.5$ and $P_+>0.5$ alternatively,
with their vertices located at the CDT points. 
We name, in short, the former pattern 
of LZS interferometry as ``symmetric resonances" (SR) and the 
latter as ``antisymmetric resonances" (AR).

\begin{figure}[ht]
\begin{center}
\includegraphics[width=0.9\linewidth,clip]{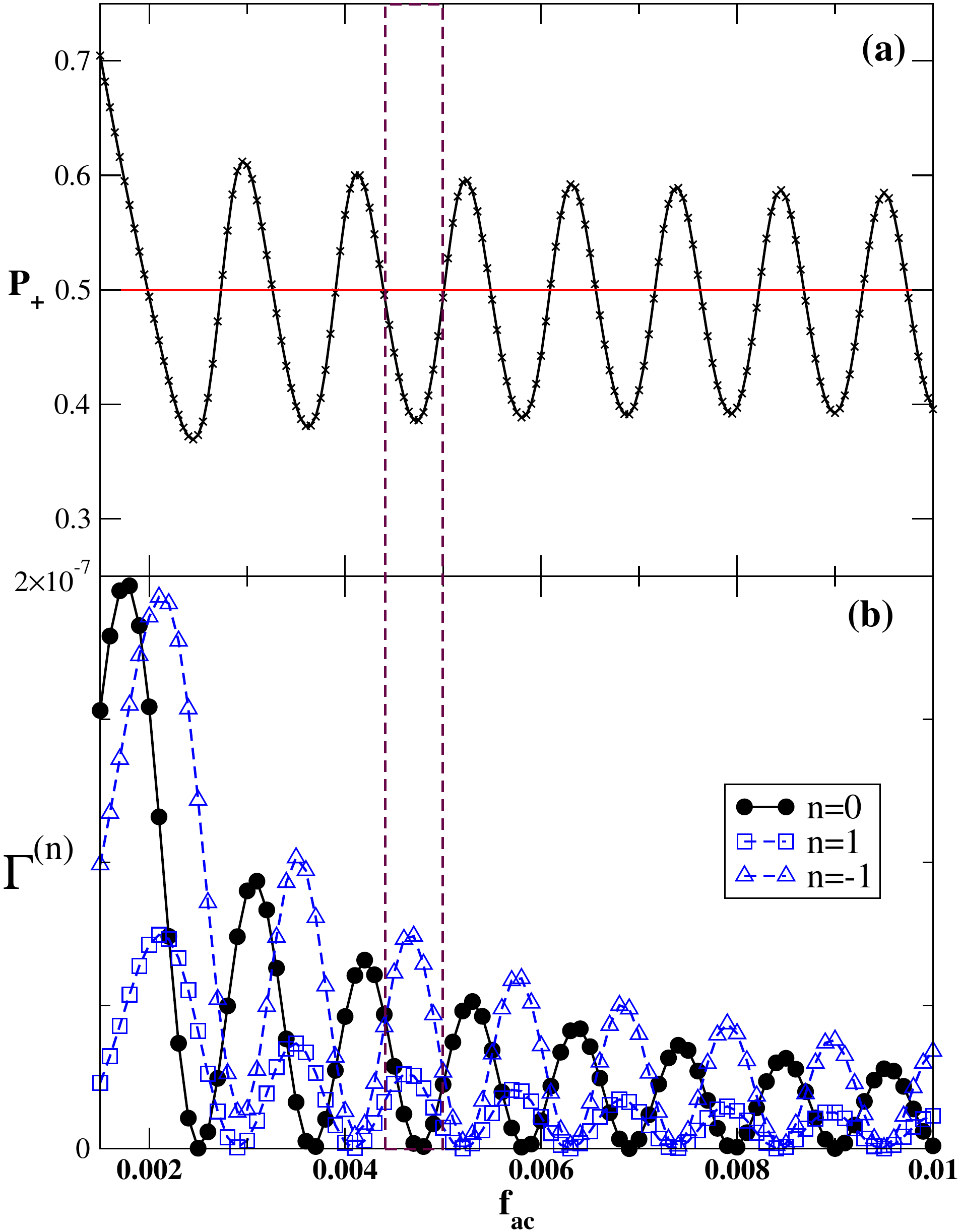}
\end{center}
\caption{(a)
Terms $\Gamma^{(n)}$ that contribute to the relaxation rate.
as a function of the driving amplitude $f_{ac}$.
The calculations were performed for $\delta f_{dc}=0.00151$,
$\omega_{0}= 2\pi/\tau= 0.003 E_J/\hbar$ and
an ohmic bath at $T=20 \rm{mK}$ (for $E_J/h\approx300{\rm GHz}$).
(b)  Asymptotic
average population $\overline P_+$ for the same case.
The dashed vertical lines highlight one of the regions of $f_{ac}$ where
there is population inversion.
} \label{fo4} 
\end{figure}

\subsection{Relaxation, decoherence and the  bath-mediated population inversion  mechanism}

The asymptotic AR interference patterns show
 population inversion (PI) on
one side of a  multiphoton resonance, as seen in Fig.\ref{fPf0}(b)
and Figs.\ref{figtemp}(c) and (d).
This makes $\overline{P_+}$ antisymmetric 
around the resonance. 
If we fix $\delta f_{dc}$ 
at a  value in between two resonances, 
the probability  oscillates around $\overline{P_+}=1/2$  as function of $f_{ac}$, 
showing PI whenever a  ``triangle'' with $\overline{P_+}<1/2 $  
is traversed. This is shown in Fig.\ref{fo4}(a).

The  underlying mechanism of this population inversion can be understood 
by  analyzing the contribution to  relaxation of virtual photon exchange 
processes with the bath.\cite{stace2005,fds2}
As we show in the Appendix, within the first diamond regime, the total
relaxation  rate $\Gamma_r$  can be decomposed as  
$$\Gamma_r = \Gamma^{(0)} +\sum_{n\not=0} \Gamma^{(n)}\;,$$
with  $\Gamma^{(n)}$  the relaxation rates due to
virtual  $n$-photon  transitions to bath oscillator
states.\cite{grifonih,stace2005,fds2}
In  Fig.\ref{fo4}(b) we plot $\Gamma^{(n)}$ as a function of $f_{ac}$ for
the same $\delta f_{dc}$ considered in Fig.\ref{fo4}(a). We  show
the cases with $n=0,\pm1$, where $\Gamma^{(0)}$ describes the
relaxation without exchange of virtual photons, corresponding to
the ``conventional'' dc relaxation mechanism, while $\Gamma^{(\pm1)}$
correspond to the the ac contribution due to
the  exchange of one  virtual
photon  with energy $\pm\hbar\omega_0$. 
We show in Fig.\ref{fo4} that, whenever there is population inversion,
 the dc relaxation terms vanish ($ \Gamma^{(n=0)}\approx 0$), 
 while the $\Gamma^{(n=-1)}$ term
  is   the largest one.  This
indicates that the relevant mechanism leading  to  PI
is a transition to a virtual level at energy 
 $E_0+\hbar\omega_{0} > E_{1}$ (one  photon absorption, $n=-1$),
followed by a relaxation to the level $E_{1}$.\cite{stace2005,fds2}

\begin{figure}
\begin{center}
\includegraphics[width=0.9\linewidth,clip]{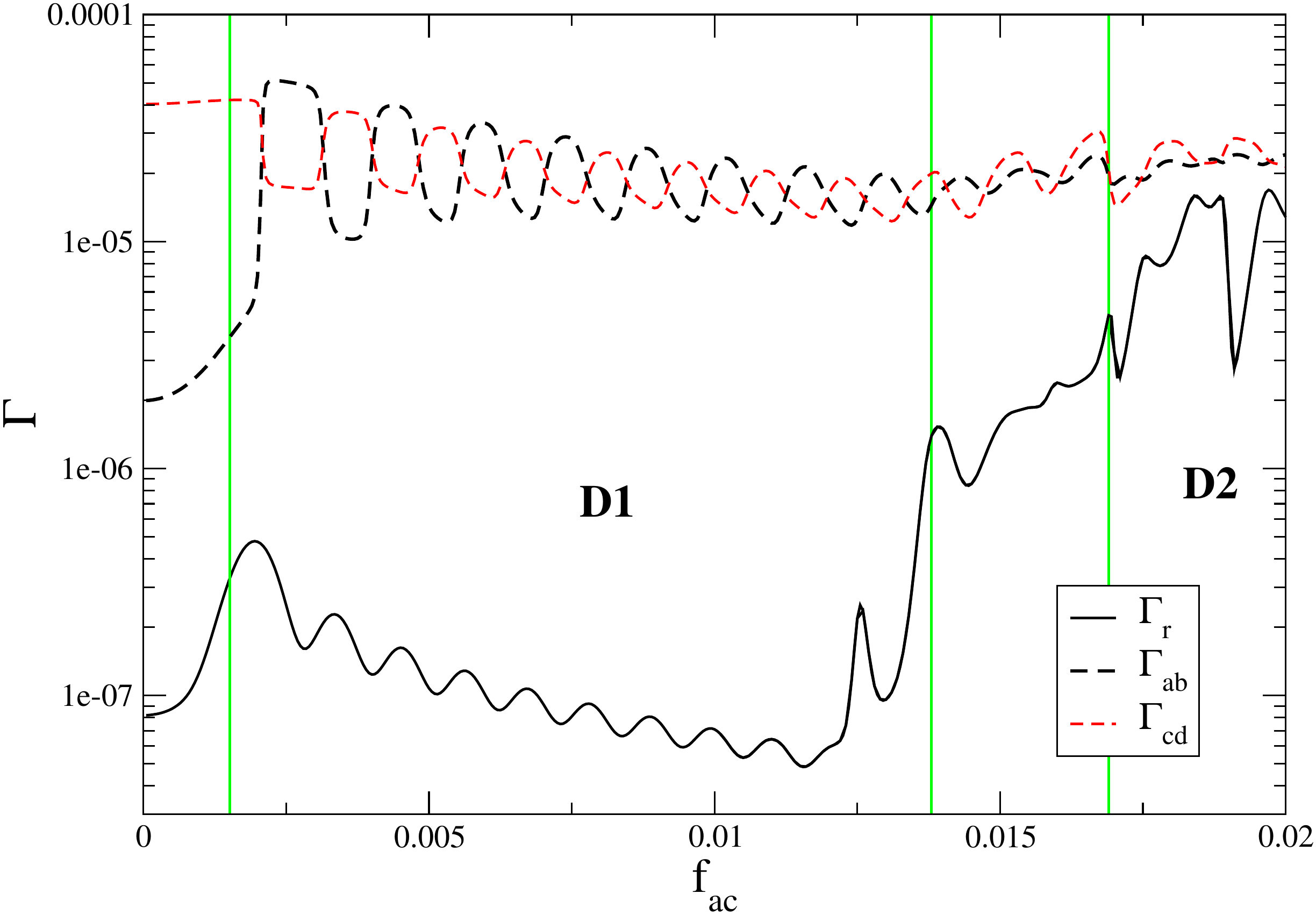}
\end{center}
\caption{
  Relaxation rate $\Gamma_{r}$ and
  decoherence rates $\Gamma_{ab}$  and $\Gamma_{cd}$ as a function of the driving amplitude $f_{ac}$.
The calculations were performed for $\delta f_{dc}=0.00151$,
$\omega_{0}= 2\pi/\tau= 0.003 E_J/\hbar$ and
an ohmic bath  at $T=20 \rm{mK}$ (for $E_J/h\approx300{\rm GHz}$).
Vertical lines separate diamond regimes D1 and D2 described in the text.
} \label{fo4c}
\end{figure}

To better understand why the AR patterns have not been observed yet
in current experiments, we now analyze the time scales of relaxation
and decoherence.  To this end, 
we calculate numerically the full relaxation rate $\Gamma_r$ and the decoherence rates $\Gamma_{\alpha\beta}$, from
the eigenvalues of the $\Lambda_{\alpha\beta\alpha'\beta'}$ matrix, as discussed in the Appendix.
In Fig.\ref{fo4c} we show the relaxation rate $\Gamma_r$ and two decoherence rates $\Gamma_{\alpha\beta}$,
as a function of the driving amplitude $f_{ac}$  for $\delta f_{dc}=0.00151$
(away from a $n$-resonance). For  $f_{ac}=0$  the relaxation rate  $\Gamma_r$   corresponds 
to the $1/T_1$ measured experimentally,
{\it i.e.},
$\Gamma_r=1/t_r\rightarrow\Gamma_1=1/T_1$
when $f_{ac}\rightarrow0$. 
Since to a good approximation the density matrix becomes diagonal
in the Floquet basis in the asymptotic regime,\cite{grifoni-hanggi} 
 the decoherence rate $\Gamma_{ab}$ shown in Fig.\ref{fo4c} 
corresponds to the decoherence between the $|a\rangle$ and
the $|b\rangle$ Floquet states.
When $f_{ac}\rightarrow0$, $\Gamma_{ab}\rightarrow\Gamma_2=1/T_2$,
since in this limit
it corresponds to the decoherence between the two lowest energy levels.
We also plot in Fig.\ref{fo4c} the rate  $\Gamma_{cd}$, which is the
decoherence rate between the $|c\rangle$ and
the $|d\rangle$ Floquet states.
At  $f_{ac}=0$ it corresponds to the decoherence between the third and the fourth energy level. 

For small $f_{ac}$,  below the onset of the first diamond  ($f_{ac}<f_{D1}^s$), the relaxation and decoherence rates stay in  values similar to
the undriven case. 
However, when $f_{ac}>f_{D1}^s$ we see in Fig.\ref{fo4c} that both  rates depend strongly on  $f_{ac}$. Above the onset of the first diamond, the overall behavior is that the
decoherence rate  $\Gamma_{ab}$  increases and the relaxation rate $\Gamma_r$ decreases as a function
of $f_{ac}$. 
Therefore, within the first spectroscopic diamond, the difference between decoherence and relaxation is much larger than in the 
undriven case ($\Gamma_{ab}\gg \Gamma_r$).
This explains the important difference between $P_+(t_{exp})$ and ${\overline P_+}$, within the first diamond
in Fig.\ref{las}, due to the  large time window
where AR patterns can be 
observed for
$\Gamma_{ab}^{-1}<t_{exp}<\Gamma_r^{-1}$.

Beyond the first diamond, for $f_{ac}>f_{D1}^e$, 
 the relaxation rate 
$\Gamma_r$ increases strongly with 
$f_{ac}$, becoming nearly of the same order
of the decoherence rates within the second diamond and above. 
This behavior 
is a consequence of the fact that
when more than  two-levels are involved in the dynamics, there
are several possible decay transitions 
between energy levels that
 contribute to a faster relaxation of the system. 
Therefore, in the second diamond
and beyond decoherence and relaxation rates become comparable. 
For this reason,
$P_+(t_{exp})\approx\overline{P_+}$ in this case,
since the relaxation time is significantly reduced
and  thus $\Gamma_{ab}^{-1}\sim\Gamma_r^{-1}<t_{exp}$.

Although previous works have found PI
for the asymptotic regime of two level systems,\cite{popin2l,hartmann,stace2005,yu2010a}
the time-dependent dynamics with different time scales 
has been overlooked.  In fact, the relevant point from our
findings is that the asymptotic regime  is difficult
to reach in the experiment, since PI  needs
long times ($t\gtrsim t_r\gg t_{dec}$) to emerge when  mediated by the bath.

\begin{figure}
	\begin{center}
		\includegraphics[width=0.8\linewidth,clip]{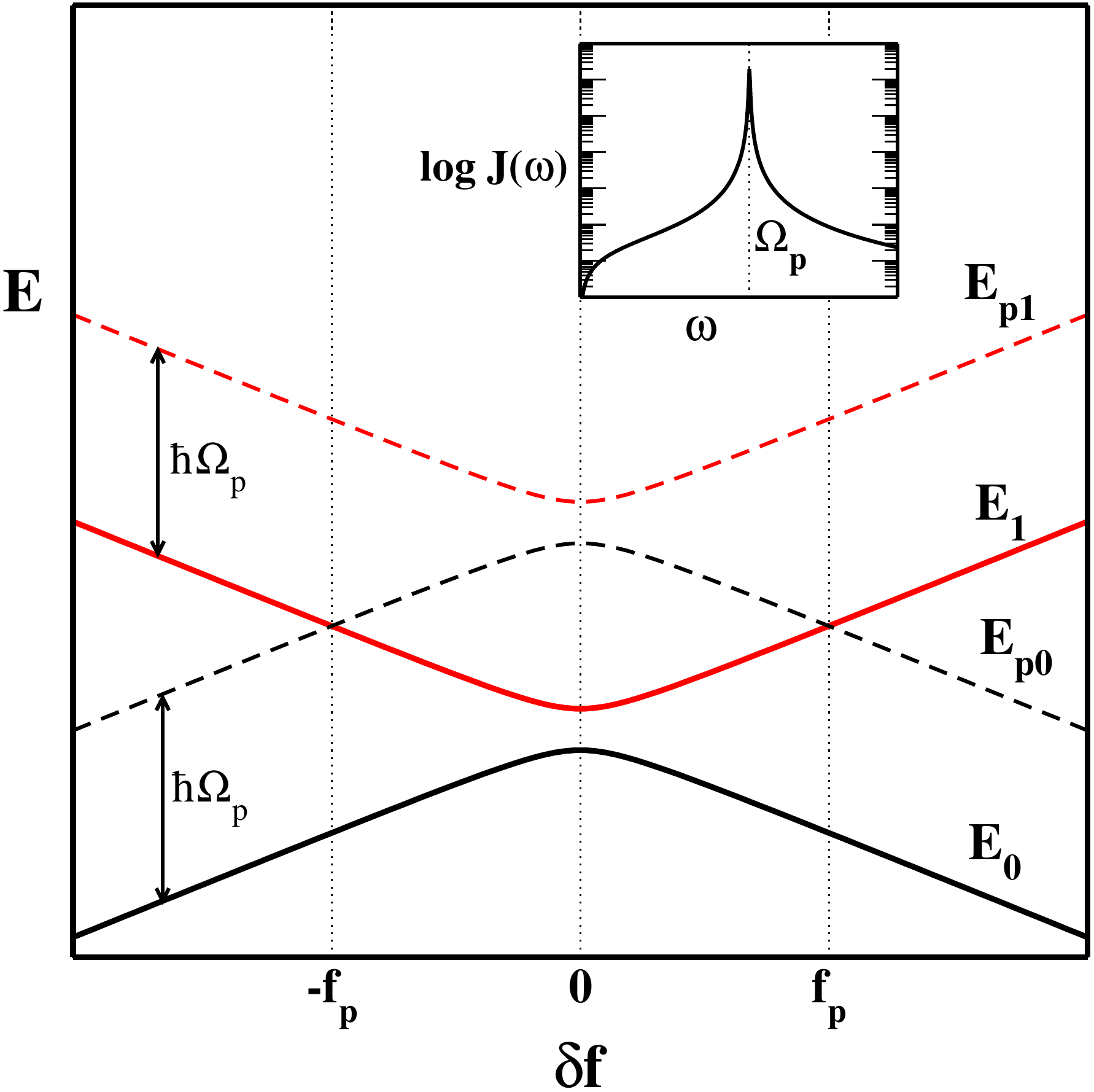}		
	\end{center}
	\caption{(Color online)	
		Schematic plot of the two low energy levels $E_0, E_1$  and the two virtual levels $E_{p0}, E_{p1}$  as a function of the flux detuning $\delta f$,  for the flux qubit coupled with an structured bath with a resonant mode at frequency $\Omega_p$. The location of the ``virtual crossings'' at $\pm f_p$ are also indicated. The inset shows a plot of the spectral density $J(\omega)$ of the structured bath.	
	} \label{majo}
\end{figure}

\subsection{Effect of a resonant mode in the  bath}

The measurement of the state of the FQ
is performed with a read-out dc SQUID, which is inductively
coupled to the qubit.\cite{oliver,valenzuela,chiorescu2,caspar}
This modifies  the bath  spectral density by adding a resonant mode  at the plasma frequency  $\Omega_p$ of the SQUID detector. 
The  so-called ``structured bath'' 
spectral density is given by\cite{caspar}
 \begin{equation}
J(\omega)=\frac{\gamma \omega\Omega_p^4}
{(\Omega_p^2-\omega^2)^2+(2\pi\kappa
\omega\Omega_p)^2},
\label{Jsb}
 \end{equation}
with $2\pi\kappa\Omega_p$ the width of the resonant peak at 
$\omega= \Omega_p$  (see the inset of
Fig.\ref{majo} for a schematic plot of $J(\omega)$).

\begin{figure}
	\begin{center}
		\includegraphics[width=0.98\linewidth,clip]{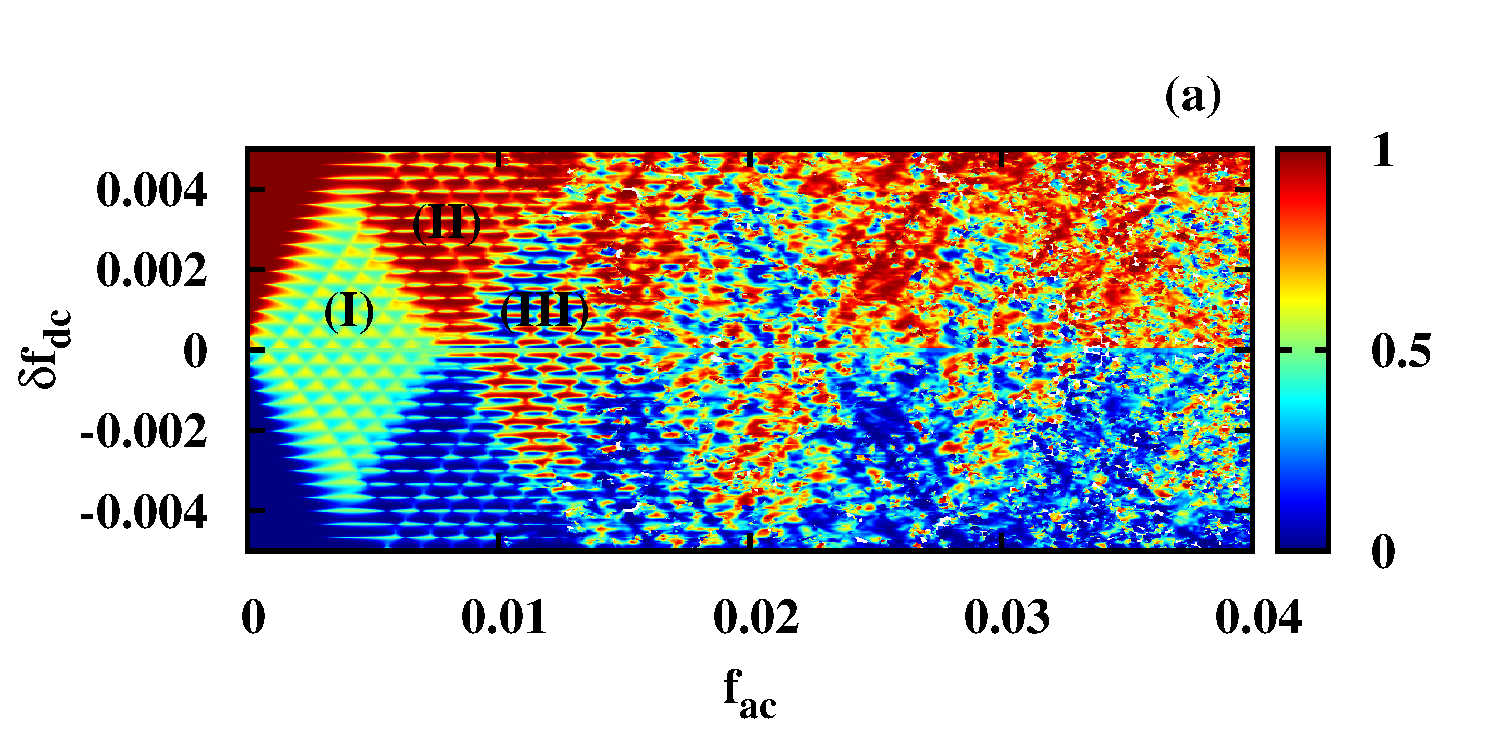}
		\includegraphics[width=0.98\linewidth,clip]{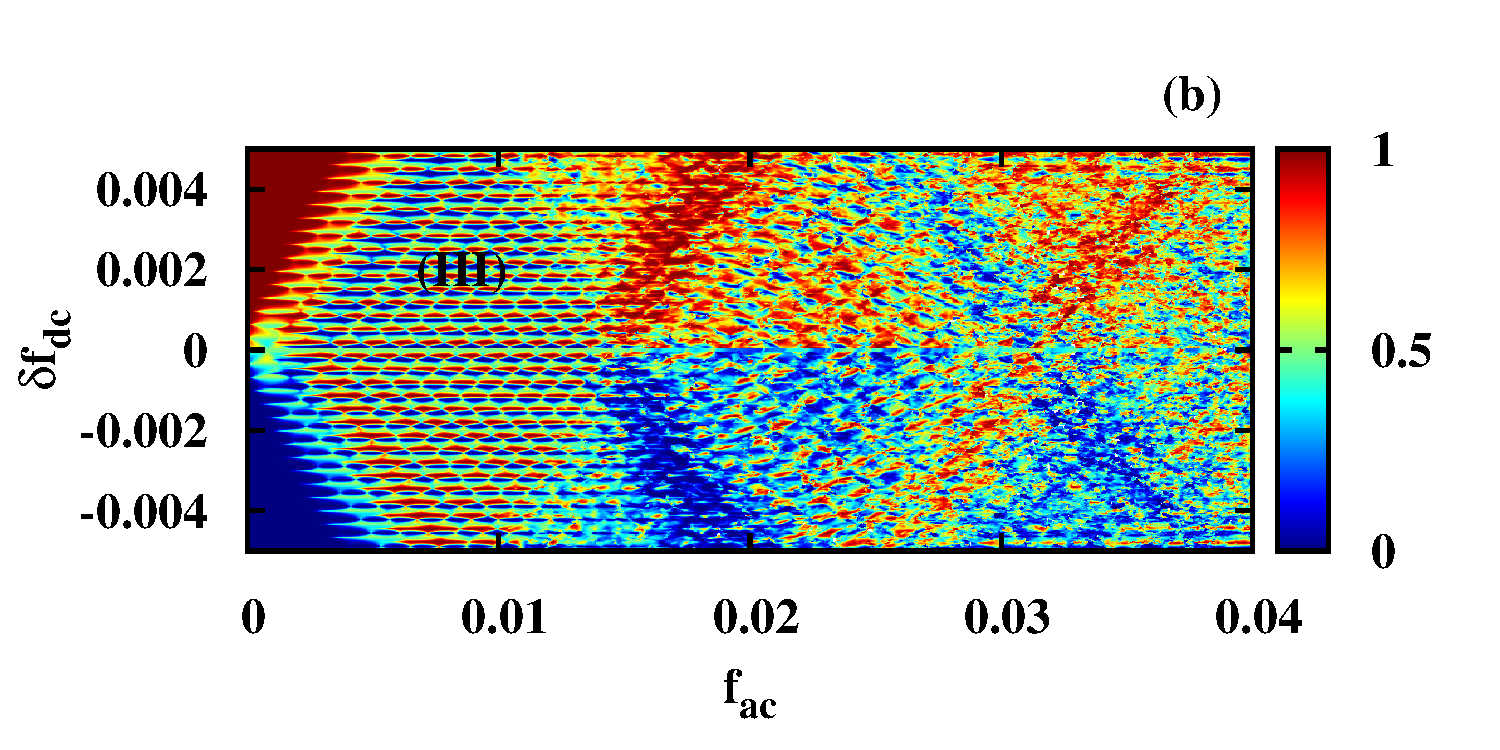}				
	\end{center}
	\caption{(Color online)
		Landau-Zener-Stuckelberg interference patterns for a flux qubit in contact with a structured bath. Showed are
		intensity plots of the asymptotic average population $\overline{P_+}$ as a function of the driving amplitude $f_{ac}$
		and dc detuning $\delta f_{dc}$ for  (a) $\Omega_p=0.08$
	    and		(b) $\Omega_p=0.02$.
	    We indicate in the plot the regions corresponding to the regimes (I), (II) and (III) described in the text.
	} \label{lasp}
\end{figure}

Here, we study the effect of this resonant mode at  $\Omega_p$ on LZS interferometry, in the case $\Omega_p>\omega_0$.
To this end, we  calculate the asymptotic ${\overline P_+}$ 
using the spectral density of Eq.(\ref{Jsb}),
considering different values of $\Omega_p$, with  $\kappa=0.001$ fixed.

In Fig.\ref{lasp} we show the intensity plots of ${\overline P_+}$ as 
a function of $(\delta f_{dc},f_{ac})$ for $\Omega_p=0.08$ and $\Omega_p=0.02$.
As it is evident, the diamond patterns are strongly affected
by the structured bath.
In the case of $\Omega_p=0.08$, we observe  in Fig.\ref{lasp}(a) that the  
$(\delta f_{dc},f_{ac})$ region formerly occupied by the first
diamond in the Ohmic case [shown in Fig.\ref{las}(b)] 
is now divided in three parts: two new sub-diamonds indicated as 
regimes (I) and (III), and the region in between them, indicated as regime (II).  
When lowering $\Omega_p$ the regime (III) becomes more predominant, 
as can be seen in Fig.\ref{lasp}(b) for $\Omega_p=0.02$.

From now on, to describe the above mentioned changes of the first diamond, 
we restrict to the two lowest levels of the FQ.
It has been shown that a two-level system coupled to an structured
bath that has a localized mode at $\Omega_p$ is
equivalent to  a two level system weakly coupled to a single mode
quantum  oscillator with frequency $\Omega_p$, and both
coupled to an Ohmic bath.\cite{garg,milenas}
In fact, most of the results discussed in this section can be qualitatively interpreted
by assuming that at low energies there are two virtual levels at $E_{p0}=E_0+\hbar \Omega_p$ 
and $E_{p1}=E_1+\hbar \Omega_p$, as sketched in Fig.\ref{majo}.
These virtual levels  are not stable and 
decay fast to their  ``underlying'' energy level,
{\it i.e.}, 
$E_{p0}\rightarrow E_0$ and $E_{p1}\rightarrow E_1$.
The three regimes found in  Fig.\ref{lasp}  can be understood by considering that
there are two ``virtual crossings'' when 
$E_1(\delta f)=E_{p0}(\delta f)$, at the
field detunings 
$$\delta f = \pm f_p \approx \pm\Omega_p/4\pi I_p,$$ 
as shown schematically in Fig.\ref{majo}.
The boundaries of these sub-diamonds can be defined 
in a similar way as  
in Sec.III.B,  replacing $f_{01}$ by $f_p$ in the argument (when $f_p<f_{01}$).
This gives that
the sub-diamond of regime (I) is
within the limits $f_{D1}^{s}< f_{ac}<f_p-\delta f_{dc}$,
and the sub-diamond of regime (III)  is within the limits
$\delta f_{dc}+f_p<f_{ac}<f_{D1}^{e}$ 
(assuming $0<\delta f_{dc}< f_p$).
However, since
the virtual levels $E_{p0}$ and $E_{p1}$ are
not truly stable, the regimes (II) and (III) show
interference patterns that are different than the analyzed
in the previous section, as we describe below.

Regime (I): {\it Antisymmetric resonances}. 
If  the nearest (in energy scale) virtual bath mode 
at $E_{p0}$ is never reached within the driving interval $\delta f_{dc}\pm f_{ac}$,
the behavior is the same as  the one analyzed previously for the Ohmic bath in Figs.\ref{las}(b) and \ref{las2}(b).
The only difference is that the AR interference pattern
is now within a smaller sub-diamond, since it is limited 
by the condition $E_1-E_0<\Omega_p$ within the driving interval.

\begin{figure}
	\begin{center}
		\includegraphics[width=0.98\linewidth,clip]{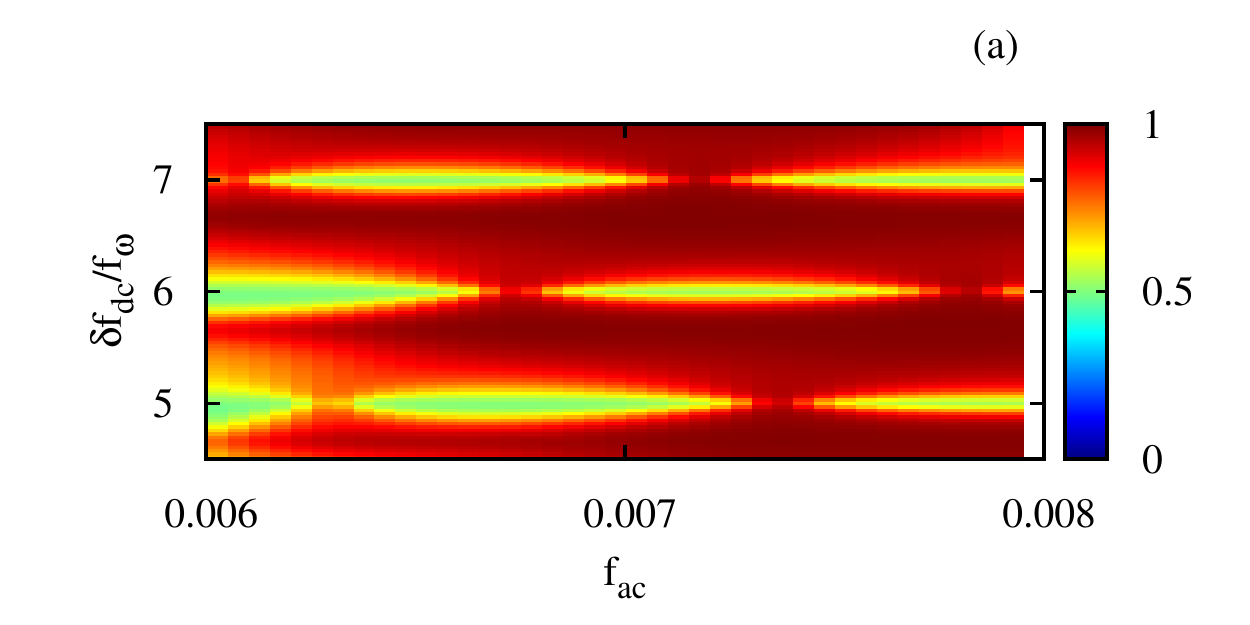}
		\includegraphics[width=0.98\linewidth,clip]{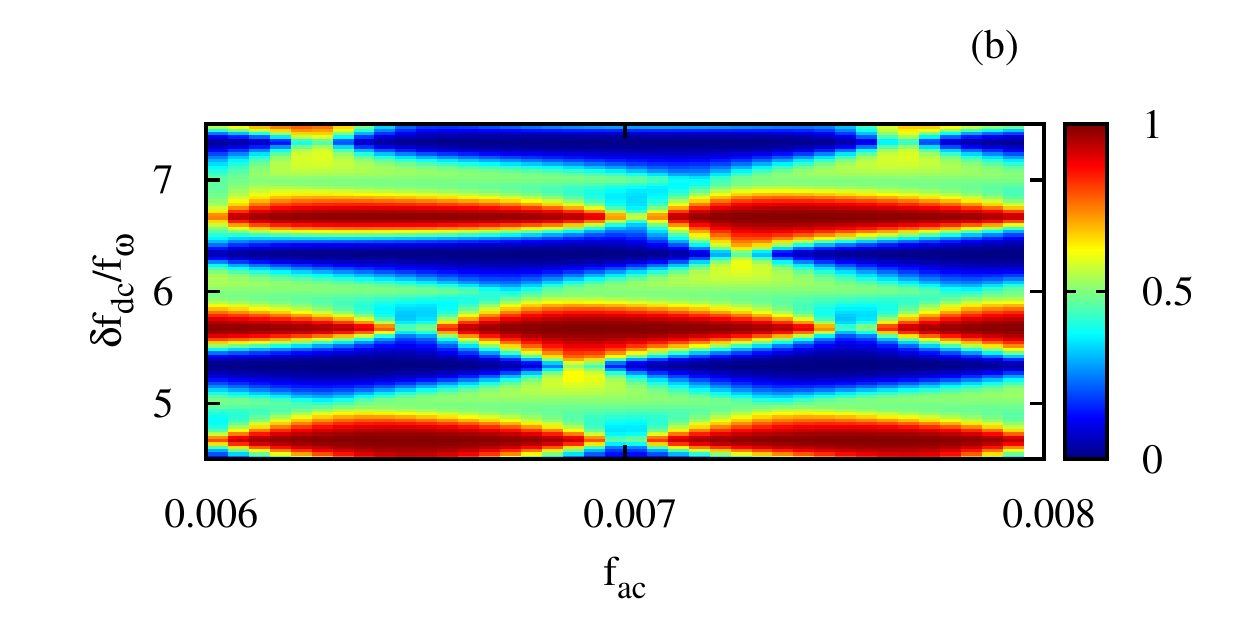}
	\end{center}
	\caption{(Color online)
	An enlarged view of  the same  section  of the first diamond
	shown in Fig.\ref{las2} for the Ohmic bath.
		(a) For $\Omega_p=0.08$. 
		(b) For $\Omega_p=0.02$
	} \label{lasp2}
\end{figure}

Regime (II): {\it Symmetric resonances}. 
In this case, one of the virtual crossings of $E_1$ with
$E_{p0}$ is reached within the 
driving interval $\delta f_{dc}\pm f_{ac}$. Therefore,
transitions from the $E_1$ level to the virtual level at $E_{p0}$ are possible.
Since this later virtual level is unstable, 
the system decays to the ground state at $E_0$. 
By going repeatedly through this process during the periodic driving,
population is pumped from the $E_1$ level to the 
ground state at $E_0$, providing
 a `cooling' mechanism. 
The transitions via the unstable mode
at $E_{p0}$ lead to a fast relaxation of the
system, before the slow mechanisms of bath mediated
population inversion, discussed in Sec.IIIC, can take place. This impedes
the dynamic transition to antisymmetric resonances.
In this way, 
the interference pattern with symmetric resonant lobes 
remains {\it in the  asymptotic regime}.
This is shown in Fig.\ref{lasp2}(a), which
corresponds to an enlarged section of the
regime (II)  of Fig.\ref{lasp}(a).

Regime (III):  {\it Sideband resonances}.
In this case, the two virtual crossings of 
$E_{1}$ with  $E_{p0}$ are reached within the driving interval
$\delta f_{dc}\pm f_{ac}$. This
makes possible to access also the $E_{p1}$ level through
Landau-Zener transitions at $\delta f=0$, which
allows for new ``sideband'' resonances involving 
the  $E_{p0}$ and $E_{p1}$ levels, in addition to the
direct resonances at $E_{1}-E_0=n\omega_{0}$.
When $E_{p1}-E_0=n\omega_0$, called a blue sideband 
resonance,\cite{milenas} the qubit resonates between
the $E_0$  and the $E_{p1}$ levels.
However, 
the $E_{p1}$ virtual level is unstable and it decays  to the $E_1$ level.
In this way, the ac drive is continuously pumping population from the ground state to
the excited state at $E_1$,
leading to full inversion of the qubit population: ${\overline P}_+\sim0$.
When $E_{1}-E_{p0}=n\omega_0$, called a red sideband resonance,\cite{milenas} 
the qubit resonates between
the $E_1$ and the $E_{p0}$ levels.
In this case, the $E_{p0}$ virtual level decays
fast to the $E_0$ level, and thus
the ac drive is continuously pumping population from the $E_1$ level
to the ground state, leading to ${\overline P}_+\sim 1$.
In Fig.\ref{lasp2}(b)  we can see in detail the sideband resonance patterns that
characterize this regime, with  alternating $P_+\sim0$ and $P_+\sim1$.
This type of resonances 
have been analyzed in Ref.\onlinecite{milenas},  
within a perturbative approach for  $\Omega_p<\omega_{0}$. 
In that case, only the regime (III) is realized. 
On the other hand,
when $\Omega_p > \omega_{0}$
the  three regimes described above are possible.

To observe the subdivision of the first diamond in the three regimes, 
the  crossing of $E_1$
with  $E_{p0}$ should occur before the avoided
crossing of $E_1$ with $E_2$. This means that the condition
 $f_p<f_{01}$ is required, or  $\Omega_p < 4\pi I_p f_{01}$.
For the  typical flux qubit parameters considered in this work,
this later condition corresponds to $\Omega_p\lesssim0.15$.
The SQUID detectors used in the measurements of Refs.\onlinecite{berns1,oliver,valenzuela} have
typically $\Omega_p \sim 10\rm{GHz}$, which is $\Omega_p \sim 0.2$ in our normalized units.
Therefore in the case of these LZS interferometry experiments,  
the effects of the resonant mode at $\Omega_p$ are negligible, 
and everything is within the regime (I) of asymptotic
asymmetric resonances.
On the opposite side, the flux qubits studied in Ref.\onlinecite{chiorescu2} have $\Omega_p \lesssim \Delta\lesssim\omega_0$, 
which situates them deeply in the 
case of the regime (III). In fact, the $n=1$ blue and red sideband resonances have been observed in Ref.\onlinecite{chiorescu2}.
However, in these devices the
oscillator mode of the SQUID detector is strongly coupled to
the qubit, while 
Eq.~(\ref{Jsb}) is valid for weak coupling.
A full analysis  in this case has to consider a
quantum  oscillator with frequency $\Omega_p$ coupled
to the qubit within the system  hamiltonian, see Refs.\onlinecite{chiorescu2,milenas}. In any case,
our results show that it will be interesting to perform
experiments on LZS interferometry using SQUID detectors
with low $\Omega_p$ and weakly coupled to the qubit, to observe
the three regimes shown in Figs.\ref{lasp} and \ref{lasp2}.

\section{Conclusions}

To summarize, by performing a realistic modeling of the 
flux qubit we were able to analyze the time dependence of
the LZS interference patterns
 (as a function of ac amplitude and dc detuning) taking into account decoherence
and relaxation. 

We found an important difference between
the LZS patterns observed for the time scale of
current experiments and those for the asymptotic
long time limit: a symmetry change in their structure. 
This is a dynamic transition as a function of time from a LZS pattern 
with nearly symmetric multiphoton resonance lobes
to antisymmetric multiphoton  resonances. 
This transition is observable only when driving the
system for very long times,
after full relaxation with the bath degrees of freedom ($t\gg t_r$).

The large time scale separation, $t_r \gg t_{dec}$, present
in the device of Ref.\onlinecite{valenzuela} explains why in their case
the asymptotic  LZS pattern is beyond the experimental  time window.
It will be interesting if experiments could be carried out
for longer driving times in this device.  
For example, measurements 
of curves of $P_+$ at growing time scales near a multiphoton resonance, 
could show the transition from symmetric to antisymmetric behavior.

Another interesting finding is the dependence of the
LZS interference patterns on the frequency $\Omega_p$ of the
SQUID detector. Different
types of LZS interference patterns can arise, depending
on the magnitude of $\Omega_p$.
In particular, we showed that the  resonant mode at $\Omega_p$ 
can impede the  dynamic transition when $\Omega_p$  is
of the order of the qubit gap, 
in the regime (II) discussed in Sec.IIID. 
In principle, the frequency $\Omega_p$ can
be varied (in a small range) by varying the driving
current of the SQUID detector \cite{chiorescu2} or with a variable shunt capacitor.

\begin{acknowledgements}

We acknowledge discussions with W. Oliver and S. Kohler.
We also acknowledge financial support from CNEA, UNCuyo (P 06/C455) CONICET
(PIP11220080101821,PIP11220090100051,PIP 11220110100981,11220150100218) and 
ANPCyT (PICT2011-1537, PICT2014-1382).

\end{acknowledgements}

\appendix*

\section{Floquet-Markov formalism for the periodically driven flux qubit}
\label{sflo}

\subsection{Periodically driven isolated flux qubit}

Since the FQ of Eq.(\ref{ham-sys}) 
is driven with a  magnetic flux 
$f(t)= f_{dc}+f_{ac}\sin(\omega_{0} t)$, the hamiltonian is time periodic ${\cal H}(t) = {\cal H}(t + \tau)$, with $\tau=2\pi/\omega_{0}$. 
In this case, it is convenient to use
the Floquet formalism, 
that allows to treat  periodic forces
of arbitrary strength and frequency. \cite{shirley,grifoni-hanggi,grifonih,fds,chu-floquet,barnes2013}
According to the Floquet theorem,\cite{shirley}
the solutions of the    Schr\"odinger equation are of the
form  
$$|\Psi(t)\rangle=\sum_{\alpha} c_\alpha e^{-i\varepsilon_\alpha t/\hbar}|\alpha(t)\rangle,$$ 
where $c_\alpha$ are time independent coefficients and
$\varepsilon_\alpha$ are the so-called quasienergies.
The  Floquet states $|\alpha(t)\rangle$  are time-periodic, 
$$|\alpha(t)\rangle=|\alpha(t+ \tau)\rangle$$
and satisfy  the eigenvalue equation
$$\hat{H}_F(t) |\alpha(t)\rangle= \varepsilon_\alpha |\alpha(t)\rangle,$$
where
$\hat{H}_F(t)={\cal H} (t)- i \hbar \partial/\partial t $ is defined
as the Floquet hamiltonian. 
The time periodicity of ${\cal H}(t)$ and $|\alpha(t)\rangle$
makes convenient to use the Fourier representation,
\begin{eqnarray}
{\cal H}(t)&=&\sum_k {\cal H}_k e^{-ik\omega_{0} t} \\
|\alpha(t)\rangle&=&\sum_k |\alpha_{k} \rangle e^{-ik\omega_{0} t}
\end{eqnarray} 
Therefore,
we can rewrite the Floquet eigenvalue equation as:
\begin{equation}
 \sum_q\left[ {\cal H}_{k-q}-\hbar k\omega_0 I \delta_{k,q} \right] |\alpha_{q}\rangle=
\varepsilon_\alpha |\alpha_{k} \rangle
\label{eqfloq1}
\end{equation}
with $I$ the identity.

For the driven FQ, we  write
${\cal H} (t)=
\ {\cal H}_{0}+ W(t)$ 
with ${\cal H}_{0}={\cal H}_{FQ}(f_{dc})$
 the time independent part of  the hamiltonian of Eq.(\ref{ham-sys}), and
\begin{eqnarray}
W(t)&=&\alpha E_J\sin[2\pi f_{ac}\sin{(\omega_{0} t)}]
\sin(2\pi f_{dc} + 2\varphi_l)+\nonumber\\
&&\!\!\!2\alpha E_J\sin^2[\pi f_{ac}\sin{(\omega_{0} t)}]\cos(2\pi f_{dc}
+2\varphi_l)\;.\nonumber\\ \label{vt1}
\end{eqnarray}
In the energy eigenbasis of ${\cal H}_{0}$, given by
${\cal H}_{0}|n\rangle=E_n|n\rangle$, the Eq.(\ref{eqfloq1}) can
be written as
\begin{equation}
\sum_{m,q}\left[(E_n-\hbar k\omega_0)\delta_{kq}\delta_{nm}+W_{k-q}^{nm}\right]\langle m|\alpha_q\rangle=
\varepsilon_\alpha \langle n|\alpha_k\rangle
\label{floq-eig}
\end{equation}

with
\begin{equation}
W^{nm}_{k}=\frac{\omega_{0}}{2\pi}\int_{0}^{2\pi/\omega_{0}}
\langle n\left|W(t)\right|m\rangle
e^{ik\omega_{0} t}  dt \;,
\label{flo9}
\end{equation}

 The static eigenvalues $E_n$
and eigenstates $|n\rangle$ are   obtained by numerical diagonalization of  
${\cal H}_{0}$, using $2\pi$-periodic boundary conditions on 
$\vec {\bf\varphi}=(\varphi_1,\varphi_2)$
and a discretization grid of $\Delta\varphi=2\pi/M$ 
(with $M=256$).\cite{foot1}
Then, the $W^{nm}_{k}$ are evaluated from Eqs. (\ref{flo9}) and (\ref{vt1}),
where the matrix elements $\langle n|\sin{(2\pi f_{dc}+2\varphi_l)}|m\rangle$ and $\langle n|\cos{(2\pi f_{dc}+2\varphi_l)}|m\rangle$ 
have been calculated using the  obtained eigenstates $|n\rangle$.
In order to solve the Floquet eigenvalue problem numerically we have to
truncate the Eq.~(\ref{floq-eig}) both in the Fourier indices $k,q$ and the in the number  of energy levels of ${\cal H}_0$ considered.\cite{fds} The truncated
eigenproblem  is  of dimension $N_d =(2K+1)N_l$ where $K= max|q-k|$ is
defined by the maximum value of the Fourier index and $N_l$  by
the number of levels considered in  the diagonalization of ${\cal H}_0$.
The obtained Floquet states $\langle n|\alpha_k\rangle$ and quasienergies 
$\varepsilon_\alpha$ contain all the information to study
the quantum dynamics of the system described above.

An alternative method, which is more efficient for
large $N_l$,  is to consider  the time evolution operator $U(t_2,t_1)$,
which in the Floquet representation can be expanded as
$$U(t_2,t_1)=\sum_{\alpha}  e^{-i\epsilon_\alpha(t_2-t_1)}|\alpha(t_2)\rangle\langle\alpha(t_1)|.$$ 
Since $|\alpha(t+ \tau)\rangle=|\alpha(t)\rangle$, the Floquet state $|\alpha(t)\rangle$ is an eigenvector
of $U(t+\tau,t)$ with eigenvalue $e^{-i\varepsilon_\alpha\tau}$. Therefore,
it is possible to calculate the Floquet states and quasienergies from
the  diagonalization of  $U(t+\tau,t)$.
Numerically, one needs to compute 
the evolution operator $U(t,0)$ within a period, for 
$0\le t\le\tau$. Taking discretized time steps
of length $\delta t = \tau/M_\tau$, we use the second-order
Trotter-Suzuki approximation
$$U(t_{j+1},t_j)\approx e^{-i {\cal H}_0\frac{\delta t}{2}} e^{-i W(t_j+\frac{\delta t}{2})\delta t}
e^{-i {\cal H}_0\frac{\delta t}{2}} $$ 
for times $t_j=j\delta t$, and we compute the product 
$U(t_n,0)=\Pi_{j=0}^{n-1} U(t_{j+1},t_j)$ for $n\le M_\tau$,
starting with $U(0,0)=I$. The Floquet states are then obtained as eigenvectors
of $U(\tau,0)\equiv U(t_{M_\tau},0)$.  
We diagonalize numerically the hermitian matrix $C=i(1+U)(1-U)^{-1}$,
solving $C|\alpha(\tau)\rangle=c_\alpha|\alpha(\tau)\rangle$,
where $c_\alpha={\rm cotan}(\varepsilon_\alpha \tau / 2)$ and $|\alpha(0)\rangle=|\alpha(\tau)\rangle$, 
The Floquet states at any time are then calculated as 
$|\alpha(t_n)\rangle = e^{i\varepsilon_\alpha t_n}U(t_n,0)|\alpha(0)\rangle$,
and their Fourier components $|\alpha_k\rangle $ can be obtained
using a fast Fourier transform routine. We find that
for $N_l\ge 4$ this numerical  procedure  is  more efficient than the 
direct diagonalization  of Eq.(\ref{floq-eig}).

Experimentally, the probability of having a state of positive or negative
persistent current in the flux qubit is measured.\cite{chiorescu,oliver,valenzuela} 
The probability of a positive current measurement (``right'' side of the 
double-well potential) can be calculated, for $\delta f \ll 1$, integrating
the probability $|\Psi(\varphi_l,\varphi_t)|^2$  within the subspace
with $\pi>\varphi_l>0$:\cite{fds}
$$P_+(t)=\int_0^{\pi}d\varphi_l \int_{-\pi}^{\pi}d\varphi_t \;|\Psi(\vec\varphi)|^2,$$
where $\vec{\varphi}=(\varphi_l,\varphi_t)$ and $\Psi(\vec\varphi)=\langle\vec\varphi|\Psi\rangle$. 
For later generalizations, it is better
to define the projector corresponding to a positive current measurement:
$$\hat{\Pi}_+=\int_{\pi>\varphi_l>0}d\vec{\varphi}\;  |\vec{\varphi}\rangle\langle\vec{\varphi}|, $$
in terms of this operator, we have $P_+(t)=\langle\Psi(t)| \hat{\Pi}_+|\Psi(t) \rangle\equiv\langle\hat{\Pi}_+ \rangle$.
For an initial condition $|\Psi_0\rangle$ at $t_0$, we can express $P_+(t)$  in the Floquet basis as
$$P_+(t)=\sum_{\alpha,\beta}e^{-i(\varepsilon_\alpha-\varepsilon_\beta)(t-t_0)}
\langle\alpha(t_0)|\Psi_0\rangle\langle\Psi_0|\beta(t_0)\rangle
\pi_{\beta\alpha}(t)  $$
with 
$\pi_{\beta\alpha}(t)=\langle\beta(t)|\hat{\Pi}_+|\alpha(t)\rangle$.
In experiments the initial time, or equivalently the initial
phase of the driving field seen by the system in repeated realizations
of the measurement, is not well defined. Then, the
quantities of interest are the probabilities averaged
over initial times, $t_0$.\cite{shirley,fds,chu-floquet} Using the properties of the Floquet functions,
the average over the initial phase time $t_0$ gives
$$\overline{e^{i(\varepsilon_\alpha-\varepsilon_\beta)t_0}|\beta(t_0)\rangle
	\langle\alpha(t_0)|}\rightarrow\delta_{\alpha\beta}\sum_k|\alpha_k\rangle
	\langle\alpha_k|,$$ 
and we obtain,
$$P_+(t)=\sum_{\alpha} {
\pi_{\alpha\alpha}(t)
\left[\sum_k |\langle\alpha_k|\Psi_0\rangle|^2 \right]}
 $$	
It is worth noticing that averaging over the initial 
phase time of the driving field is equivalent to defining
a density matrix $\rho(t)=\overline{|\Psi(t)\rangle\langle\Psi(t)|}$
which, due to the average over $t_0$, is diagonal in the Floquet basis,
$\rho_{\alpha\beta}=
\langle\alpha(t)|\rho(t)|\beta(t)\rangle=\delta_{\alpha\beta}\sum_k |\langle\alpha_k|\Psi_0\rangle|^2$.

Finally, we average within a period over the observation time $t$,
to obtain a time independent ``stationary'' probability $\overline{P}_+\equiv \overline{P_+(t)}$. 
After defining the time-averaged projector in the Floquet basis as
$$ \overline{\pi}_{\alpha\beta}=\overline{\langle\alpha(t)|\hat{\Pi}_+|\beta(t)\rangle}
	= \sum_k\langle\alpha_k|\hat{\Pi}_+|\beta_k\rangle$$
we obtain the simple result
$$\overline{P}_+=\sum_{\alpha} \overline{\pi}_{\alpha\alpha}\rho_{\alpha\alpha}$$
with $\rho_{\alpha\alpha}=\sum_k |\langle\alpha_k|\Psi_0\rangle|^2$.		
Numerically, after diagonalization of ${\cal H}_0$ we compute
the matrix elements of the projector $\pi_{nm}=\langle n|\hat{\Pi}_+|m\rangle=
\int_{\pi>\varphi_l>0}\Phi_n^*(\vec{\varphi})\Phi_m(\vec{\varphi})d\vec{\varphi}$,
with $\Phi_n(\vec{\varphi})=\langle\vec\varphi|n\rangle $.\cite{fds}
Then, for each $\omega_0$ and $f_{ac}$ the Floquet states and quasienergies
are obtained, and the coefficients 
$\overline{\pi}_{\alpha\beta}=\sum_{k,n,m}\pi_{n,m}\alpha_{k,n}^*\beta_{k,m}$
are evaluated (with $\alpha_{k,n}=\langle n|\alpha_k\rangle$).

\begin{figure}[ht]
	\begin{center}
		\includegraphics[width=0.9\linewidth,clip]{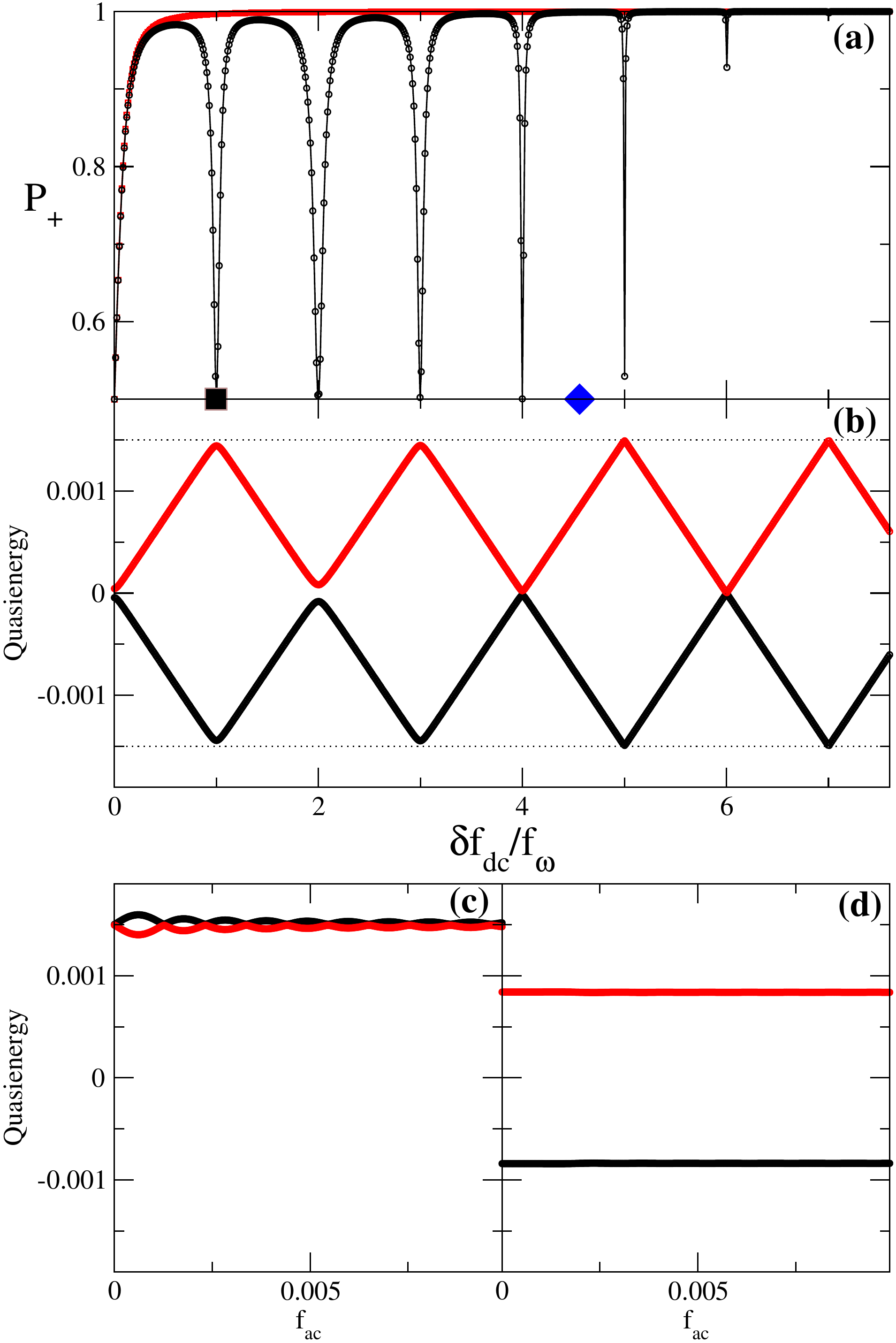}
	\end{center}
	\caption{(Color online)
		(a) Probability $\overline{P}_+$ of a state with positive loop current as 
		a function of $\delta f_{dc}$ for
		the undriven qubit (red squares) and for the
		driven qubit (black circles) with $f_{ac}=0.0001$ 
		and $\hbar\omega_{0}/E_J=0.0003$. 
		The flux detuning $\delta f_{dc}$ is normalized by $f_\omega=\omega_0/4\pi I_p$, such that the $n$-photon resonances are at $\delta f=nf_\omega$.
		(b) Floquet quasienergies (in units of $E_J$) as a function of $\delta f_{dc}$
		for the same case as in (a).
		(c) Floquet quasienergies (in units of $E_J$) as a function
		of $f_{ac}$ for the $n=1$ resonant state
		at $f_{dc}=0.50033$ [corresponding to black square in panel (a)].
		(d) Floquet quasienergies (in units of $E_J$) as a function
		of $f_{ac}$ for an out of resonance state
		at $f_{dc}=0.50151$ [corresponding to blue diamond in panel (a)].
		Device parameters of the flux qubit are $\alpha=0.8$ and $\eta=0.25$.
	} \label{qef0}
\end{figure}

As an example, we calculate ${P}_+$ for the driven FQ
in the two-level regime, as described by the hamiltonian ${\cal H}_{FQ}\approx{\cal H}_{TLS}$ of Eq.(\ref{htls}).
At zero temperature and in the absence of driving ($f_{ac}=0$) the
isolated  FQ is in the ground state. 
In this case, the probability
$P_+$ is simply the projection of the ground state on the
subspace of positive persistent current, 
and we have $P_+=\langle0|\hat\Pi_+|0\rangle=\pi_ {00}$.
 For $\delta f_{dc}=f_{dc}-1/2>0$ we have $ P_+ \approx 1$ except near
$\delta f_{dc}=0$ where $ P_+=1/2$, as can be seen in Fig.\ref{qef0}(a).
(While for $\delta f_{dc}<0$, we have $ P_+ \approx 0$, since the ground
state has the loop current in the opposite direction in this case).
In the presence of an ac drive, $\delta f(t)=\delta f_{dc}+f_{ac}\cos(\omega_{0} t)$, 
we calculate the time averaged ${\overline P}_+$,
following the procedure discussed above.
The  time averaged ${\overline P}_+$ vs. 
$f_{dc}$ shows dips corresponding to  $n-$photon resonances.    
as shown in Fig.\ref{qef0}(a).
These $n$-resonances are at $\epsilon_0 \approx n \omega_{0}$,
which is equivalent to $\delta f _{dc}=nf_\omega$, after
defining $f_\omega = \omega_0/4\pi I_p$.
the $n$-resonances are at $\delta f _{dc}=nf_\omega$.
In the Floquet picture, 
these resonances correspond to avoided crossings of the Floquet 
quasienergies\cite{grifoni-hanggi,grifonih} as a function of $\delta f_{dc}$
as we illustrate  in Fig.\ref{qef0}(b).
When increasing $f_{ac}$, we
see that for a $n$-resonance the quasienergies 
have a small gap,
$|\varepsilon_\alpha-\varepsilon_\beta| \ll \omega_{0}$ (with the difference $\varepsilon_\alpha-\varepsilon_\beta$ defined modulo $\omega_0$). On the other hand, for  a value of $\delta f_{dc}$ away of a resonance  the quasienergies 
maintain a finite gap (compared to $\omega_0$) as a function of $f_{ac}$, see Fig.\ref{qef0}(c) and (d).

\subsection{Floquet-Markov approach for  open system}

Experimentally, the system is affected by the electromagnetic
environment that  introduces decoherence and relaxation  processes.
A  standard theoretical model to study  enviromental effects  is to
couple the system bilinearly 
to a bath of non-interacting harmonic oscillators with masses $m_\nu$, 
frequencies $\omega_\nu$, momenta $p_\nu$, and coordinates $x_\nu$, with 
the coupling strength $\gamma_\nu$.\cite{grifoni-hanggi}
 The total Hamiltonian of system 
and bath is then given by 
$${\cal H} = {\cal H}_S(t) + {\cal H}_{SB}+ {\cal H}_B$$
where ${\cal H}_S(t)$ is the time-periodic  Hamiltonian of the system, $H_B$ is the Hamiltonian that describes a  bath
of harmonic oscillators  and $H_{SB}$ 
its system-bath coupling Hamiltonian term, 

\begin{equation}
\label{hb}
{\cal H_B}=\sum_{\nu=1}^{\infty}\left[\frac{p_\nu^2}{2m_\nu}+
\frac{m_\nu\omega_\nu^2 x_\nu^2}{2}\right]
\end{equation}
 
\begin{equation}
\label{hsb}
{\cal H_{SB}}=\hat{A}\sum_{\nu=1}^{\infty}\gamma_\nu x_\nu 
\end{equation}

\noindent 
with $\hat{A}$ the operator of the system that couples to the bath.
The bath degrees of freedom are characterized by the spectral density
$$J(\omega)=\pi\sum_{\nu=1}^{\infty}\frac{\gamma_\nu^2}{2m_\nu\omega_\nu}
\delta(\omega-\omega_\nu).$$ 
It is further assumed that at time $t=0$ the 
bath is in thermal equilibrium and uncorrelated to the system. Then, the full
density matrix $\sigma(t)$  has at initial time
the form $\sigma(0)=\rho(0)\exp(-\beta H_B)/tr_B\exp(-\beta H_B)$, where
$\rho(t)= {\rm Tr}_B(\sigma)$ is the density matrix of the system and $T=1/(k_B\beta)$ is the bath temperature.  After expanding the density
matrix of the system   in the time-periodic  Floquet states
\begin{equation}
\rho_{\alpha\beta}(t)=\langle \alpha(t)|\rho(t)|\beta(t)\rangle\;,
\end{equation}
the Born (weak coupling) and Markov (fast 
relaxation) approximations for the time evolution of 
$\rho_{\alpha\beta}(t)$ are performed.
In this way, the Floquet-Markov master equation  is 
obtained\cite{grifoni-hanggi,grifonih,hanggi,breuer,ketzmerich,fazio1,solinas2}
\begin{equation}
\frac{d\rho_{\alpha\beta}(t)}{dt}=-\frac{i}{\hbar}
(\varepsilon_\alpha-\varepsilon_\beta)\rho_{\alpha\beta}(t) +
\sum_{\alpha'\beta'}{\cal L}_{\alpha\beta\alpha'\beta'}(t)\rho
_{\alpha'\beta'}(t)
\label{fbr1}
\end{equation}
The first term in Eq.(\ref{fbr1}) represents the nondissipative 
dynamics and the influence of the bath is described by the
time-dependent rate coefficients
\begin{equation}
{\cal L}_{\alpha\beta\alpha'\beta'}(t)= \sum_q{\cal L}^{q}_{\alpha\beta\alpha'\beta'}e^{-iq\omega_{0} t}
\end{equation}
with
\begin{eqnarray}
&&{\cal L}^{q}_{\alpha\beta\alpha'\beta'}= \sum_{k}\left(
g_{\alpha \alpha'}^k+g_{\beta\beta'}^{-k-q}\right) A_{\alpha \alpha'}^k
A_{\beta' \beta}^{k+q}-\nonumber\\
&&\delta_{\beta \beta'}\sum_{\eta} g_{\eta \alpha'}^kA_{\alpha 
\eta}^{k+q}A_{\eta \alpha'}^k- \nonumber \\
&&\delta_{\alpha \alpha'}\sum_{\eta} g_{\eta 
\beta'}^{-k}A_{\eta\beta}^{k+q}A_{\beta' \eta}^k
\label{fbr2}
\end{eqnarray}

The nature of the bath is encoded in the 
coefficients
 $$g_{\alpha \beta}^q=J(\varepsilon_{\alpha\beta}^q/{\hbar})n_{\rm th}(\varepsilon_{\alpha\beta}^q)$$ 
with $\varepsilon_{\alpha\beta}^q=\varepsilon_\alpha-\varepsilon_{\beta}+q\hbar \omega_{0}$
and $n_{\rm th}(x)=1/(\exp{(x/k_BT)}-1)$, and
defining $J(-x)=-J(x)$ for $x<0$. The system-bath interaction is encoded
 in the transition matrix elements
 $A_{\alpha \beta}^k$  in the Floquet basis

$$A_{\alpha\beta}^q=\sum_{k}\langle\alpha_{k}|\hat{A}  |\beta_{k+q}\rangle$$

In the case of the driven FQ, the system hamiltonian is ${\cal H}_S={\cal H}_{FQ}(t)$ 
and the bath degrees of freedom couple with
the system variable $\varphi_l$ since
the dominating source of decoherence is flux noise
 (see Ref.\onlinecite{caspar}).
Thus, after  taking $\hat{A}=\varphi_l$ and in terms of the
eigenbasis of $H_0$, we have to compute
\begin{equation}
A_{\alpha\beta}^q=\sum_{nm}\sum_{k} \alpha_{k,n}^*\beta_{k+q,m}\langle n|\varphi_l|m\rangle.
\label{phim2}
\end{equation}

Considering that  the time scale $t_r$ for full relaxation satisfies $t_r\gg \tau$,
the transition rates ${\cal L}_{\alpha\beta\alpha'\beta'}(t)$ can be approximated by their average over  one period $\tau$, \cite{hanggi} 
 ${\cal L}_{\alpha\beta\alpha'\beta'}(t)
\approx{\cal L}^{q=0}_{\alpha\beta\alpha'\beta'}={\it L}_{\alpha\beta\alpha'\beta'}$, obtaining
\begin{eqnarray}
{\it L}_{\alpha\beta\alpha'\beta'}&=& R_{\alpha\beta\alpha'\beta'}
+R_{\beta\alpha\beta'\alpha'}^* \\
& &-\sum_\eta \left( \delta_{\beta \beta'}
R_{\eta\eta\alpha'\alpha}+ 
\delta_{\alpha \alpha'}R_{\eta\eta\beta'\beta}^* \right).
\nonumber
\end{eqnarray} 
The rates 
\begin{equation}
R_{\alpha\beta\alpha'\beta'} = \sum_{q} g_{\alpha \alpha'}^qA_{\alpha \alpha'}^qA_{\beta' \beta}^{-q}
\label{rates}
\end{equation}
can be interpreted as  sums of $q$-photon exchange terms.

This formalism  has been extensively
employed to study relaxation and decoherence for
time dependent periodic evolutions in double-well potentials
and in two level systems.\cite{grifoni-hanggi,grifonih,hanggi,breuer,ketzmerich}
Here we  use it to model
the ac driven FQ, considering the full multilevel Hamiltonian of Eq.~(\ref{ham-sys}).\cite{fds2}

The probability of a positive current measurement 
is calculated as 
$P_+(t)=\langle\hat\Pi_+\rangle={\rm Tr}(\hat\Pi_+ \rho(t))$, 
with $\hat\Pi_+$ the projector
defined in the previous section. 
With $\rho(t)$ calculated in the Floquet basis is 
\begin{equation}
P_+(t) = \sum_{\alpha,\beta}\pi_{\beta\alpha}(t)\rho_{\alpha\beta}(t)
\label{p+t}
\end{equation}

To calculate the time dependence of $\rho_{\alpha\beta}(t)$
it is convenient to work in the superoperator
formalism of the so-called Liouville space.\cite{blum,pollard}
We write,
$$ \frac{d\rho_{\alpha\beta}(t)}{dt} = 
\sum_{\alpha'\beta'} \Lambda_{\alpha\beta\alpha'\beta'}\;\rho_{\alpha'\beta'}
$$
with $\Lambda_{\alpha\beta\alpha'\beta'}=-\frac{i}{\hbar}
(\varepsilon_\alpha-\varepsilon_\beta)\delta_{\alpha\alpha'}\delta_{\beta\beta'}+ L_{\alpha\beta\alpha'\beta'}$.
Then we change notation rewriting the $N_l\times N_l$ matrix $\rho$ as 
an  $N_l^2\times1$ vector represented as the ket $|\rho\rangle\rangle$, 
and the $N_l^2\times N_l^2$  ``supermatrix'' $\Lambda_{\alpha\beta\alpha'\beta'}$
as the operator $\hat{\Lambda}$ acting on this linear space, where
the inner product is
defined as $\langle\langle\sigma|\rho\rangle\rangle={\rm Tr}(\sigma^\dagger\rho)$.
In particular, for the identity matrix $I$
we have $\langle\langle I|\rho\rangle\rangle=\langle\langle\rho|I\rangle\rangle=
{\rm Tr}(\rho)=1$, the later equality corresponding to 
the normalization of $\rho$, which is a conserved quantity.
On the other hand, the norm of the vector $|\rho\rangle\rangle$
is $|||\rho\rangle\rangle||=\sqrt{{\rm Tr}(\rho^2)}\le 1 $.
In this notation, we can rewrite the Floquet-Markov equation  as
\begin{equation}
	\frac{d|\rho\rangle\rangle}{dt} = \hat{\Lambda}|\rho\rangle\rangle
\label{eqFM}	
\end{equation}

The superoperator $ \hat{\Lambda}$ is non-hermitian and
has left and right eigenvectors with complex eigenvalues $\lambda_\mu$,
\begin{eqnarray}
\hat{\Lambda}|r_\mu\rangle\rangle&=&\lambda_\mu|r_\mu\rangle\rangle\\
\langle\langle l_\mu|\hat{\Lambda}&=&\langle\langle l_\mu|\lambda_\mu
\end{eqnarray}
which are mutually orthogonal, 
$\langle\langle l_\mu|r_\nu\rangle\rangle=\delta_{\mu\nu}$. 
In general, the number of independent eigenvectors of $\hat{\Lambda}$
can be less than the dimensionality of $\hat{\Lambda}$. A formal
solution of Eq.(\ref{eqFM})  for  $|\rho(t)\rangle\rangle$ can
be obtained using a similarity transformation to the Jordan normal form of $\hat{\Lambda}$.\cite{lendi,baumgartner,albert} 
In the cases considered in this work we found
numerically that it was always possible to diagonalize $\hat{\Lambda}$,
in which case the solution of Eq.(\ref{eqFM})  can be expressed as
\begin{eqnarray}
|\rho(t)\rangle\rangle&=&\sum_{\mu} c_\mu e^{\lambda_\mu t} |r_\mu\rangle\rangle\\
c_\mu&=&\langle\langle l_\mu|\rho(0)\rangle\rangle
\label{rhot1}
\end{eqnarray}
from where we can calculate numerically $\rho_{\alpha\beta}(t)\equiv |\rho(t)\rangle\rangle_{\alpha\beta}$.
The  probability of a positive current measurement is then obtained combining Eq.(\ref{p+t}) with Eq.(\ref{rhot1}).

The asymptotic state $|\rho(t\rightarrow\infty)\rangle\rangle
\equiv|\rho^\infty\rangle\rangle$ satisfies,
$$ \hat{\Lambda}|\rho^\infty\rangle\rangle =0 $$
Therefore, the asymptotic state can be constructed
from the right-eigenvectors $|r_\mu\rangle\rangle$   of $\hat{\Lambda}$ with eigenvalue $\lambda_0=0$
({\it i.e.}, the kernel of  $\hat{\Lambda}$).
If $\lambda_0$ is non-degenerate,  then the asymptotic state is  unique  and independent of
the initial condition $\rho(0)$. 
In the cases considered in this work we found,
within the numerical accuracy, that the eigenvalue $\lambda_0=0$ was non-degenerate, and so the
asymptotic  state was given by the eigenvector $|r_0\rangle\rangle$, and then $\rho_{\alpha\beta}^\infty=|r_0\rangle\rangle_{\alpha\beta}$. 
The  time independence of the asymptotic $\rho_{\alpha\beta}^\infty$,
implies that quantities  like $P_+$ in Eq(\ref{p+t}),
become time-periodic with period $\tau$ in the asymptotic state. Therefore, it is convenient to calculate the asymptotic $P_+(t)$ averaged over one period as
\begin{equation} 
\overline{P_+} = \sum_{\alpha,\beta}\overline{\pi}_{\beta\alpha }|r_0\rangle\rangle_{\alpha\beta}
\end{equation}

It is also clear from Eq.(\ref{rhot1}) that
information about the relaxation and decoherence rates is contained in the non-zero 
eigenvalues $\lambda_\mu$ of $\hat{\Lambda}$.\cite{lendi,baumgartner,albert} On one hand, the relaxation rates are given by the negative real eigenvalues of $\hat{\Lambda}$,  where the long time relaxation rate $\Gamma_r=1/t_r$ is given by  the minimum of $-Re(\lambda_\mu)$ (excluding $\lambda_0=0$). 
On the other hand,
the decoherence rates are given by  the negative real parts of the complex conjugate pairs of eigenvalues of $\hat{\Lambda}$.

\subsection{Decoherence in the Floquet basis}

It is well known that the density matrix becomes diagonal in
the energy basis for times larger than the relaxation time in undriven systems.\cite{blum} In this case, 
the decay time of the offdiagonal $\rho_{nm}$ defines
the decoherence time $\tau_{\phi}^{nm}$ between the eigenstates $E_n$ and $E_m$.
For time scales $t\gg \tau_{\phi}^{nm}$, relaxation
is described by the Pauli rate equation
for the populations of the energy levels $P_n=\rho_{nn}$.
In the case of a system with a time periodic drive, it is
usually assumed that
for large times  the density matrix  becomes 
approximately diagonal in the
Floquet basis.\cite{grifoni-hanggi,hanggi,breuer,ketzmerich,fazio1}
More precisely,
this approximation can be justified when
$\varepsilon_\alpha -\varepsilon_\beta \gg  {\it L}_{\alpha'\beta'\alpha\beta}$,
which is fulfilled for very weak coupling 
with the environment and away from resonances, see Ref.\onlinecite{hanggi,fazio1}. 
From Fig.\ref{qef0}(c) and (d) it is clear that this condition will be easily satisfied in the offresonant case considered
here, where the Floquet gap $|\varepsilon_\alpha -\varepsilon_\beta|$
is large. On the other hand at an near a resonance where
$|\varepsilon_\alpha -\varepsilon_\beta|\approx0$ this condition can not be fulfilled, unless
the system-bath coupling is extremely small.\cite{fazio1}

\begin{figure}[th]
	\begin{center}
		\includegraphics[width=0.9\linewidth,clip]{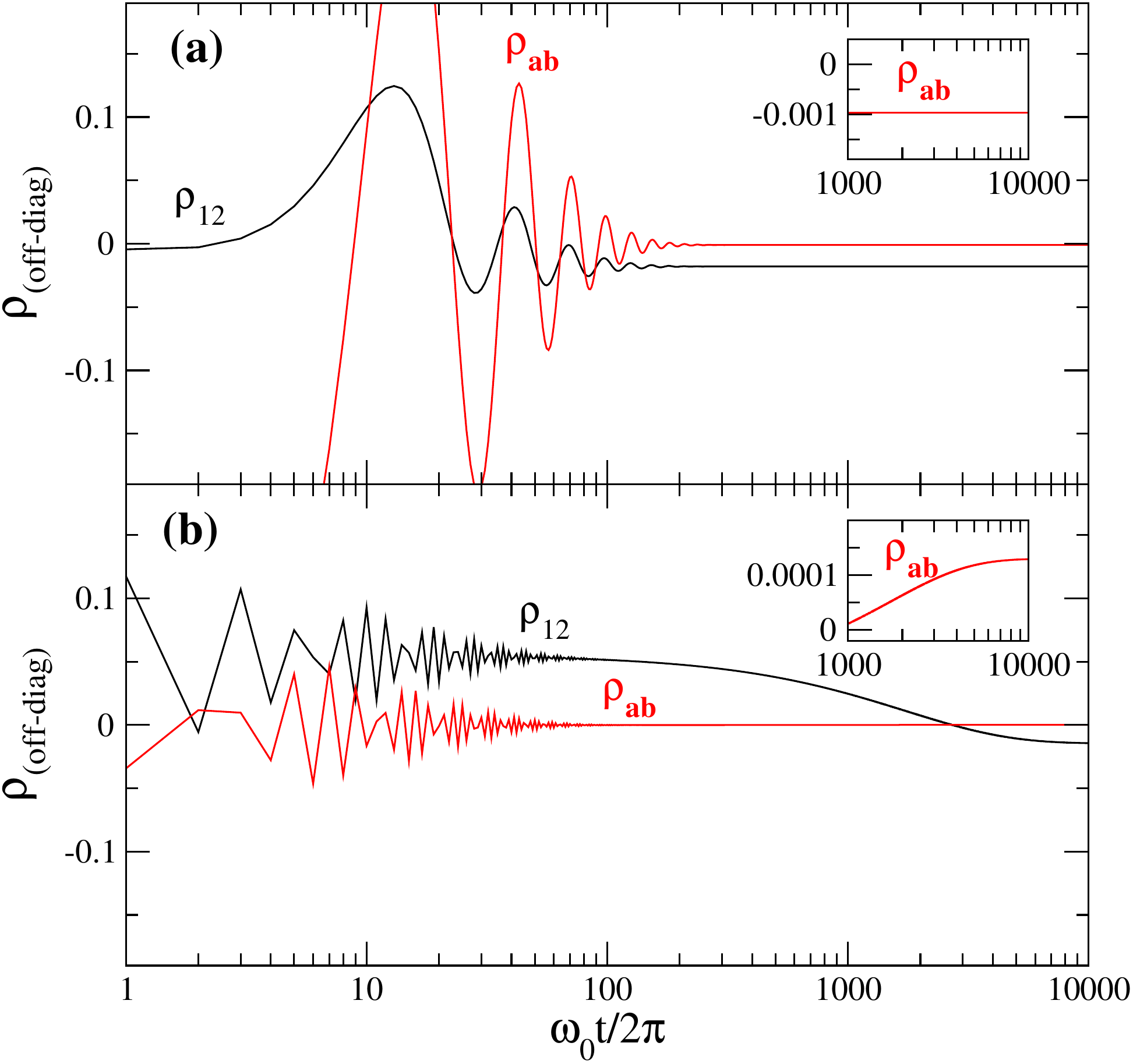}
	\end{center}
	\caption{(Color online)
		Off-diagonal elements of the density matrix $\rho$ as a function of time.
		Black lines: off-diagonal matrix element $\rho_{12}$ in the
		basis of eigenstates of the undriven Hamiltonian.
		Red lines: off-diagonal element $\rho_{ab}$ in the Floquet basis.
		(a) For a resonant state at $f_{dc}=0.50099$ driven with
		$f_{ac}=0.0001$ and $\hbar\omega_{0}/E_J=0.0003$. 
		(b) For an off-resonant state at $f_{dc}=0.50151$ driven with
		$f_{ac}=0.00245$ and $\hbar\omega_{0}/E_J=0.0003$. 
		Insets in (a) and (b)
		show $\rho_{ab}$ at large times.
	} \label{foff}
\end{figure}

In Fig.\ref{foff} we show the time
evolution of  matrix elements of $\rho$ 
calculated in the eigenbasis of $H_0$ and in the Floquet basis, 
both for an offresonant and for a  resonant case.
To clarify notation, for the $H_0$ eigenbasis we use latin indices $i,j$ with $i=1,2,3,\ldots$ ordered 
for increasing eigenenergy $E_1 < E_2 < E_3 < \ldots$. 
For the Floquet basis we use greek indices $\alpha,\beta$
with $\alpha=a,b,c,\ldots$, and ordered such that  $\alpha=a$
corresponds to the state that in the limit $f_{ac}\rightarrow0$ maps to the ground state (with index $i=1$);  
$\alpha=b$
corresponds to the state that in the limit $f_{ac}\rightarrow0$ maps
to the first excited state (with index $i=2$), and so on.

The time dependence of the  offdiagonal elements
$\rho_{ij}$ in the the $H_0$ eigenbasis, are shown
in Fig.\ref{foff}(a) and Fig.\ref{foff}(b) for the resonant and
offresonant cases, respectively.
(Here we consider the flux qubit in contact with an Ohmic bath
with $J(\omega)=\gamma\omega$).
 We see that for long times $\rho_{ij}(t\rightarrow\infty)$
goes to  a finite non-zero value that can not be neglected,
showing explicitly that the driven system density matrix
is not diagonal in the eigenenergy basis. 
Therefore approaches based on the use of the 
Pauli rate equation in the eigenenergy
basis will not be correct for the analysis of the asymptotic
long time behavior.
In the case of the Floquet basis, 
we see in Fig.\ref{foff}(a) and (b) 
that the offdiagonal $\rho_{ab}$  after
having oscillations at short time scales, decreases exponentially
to very low values for long times. We find that  
$\rho_{ab}(t\rightarrow\infty)\approx 10^{-3}$ for the
offresonant case  (while diagonal elements 
$\rho_{\alpha\alpha}(t\rightarrow\infty)\sim 0.1 - 1$).
This confirms that it is a good approximation to neglect the 
$\rho_{ab}$ at long times in this case. On the other hand,
in the resonant case 
we find $\rho_{ab}^{\rm resonant}(t\rightarrow\infty)\approx 10^{-2} >
\rho_{ab}^{\rm offresonant}(t\rightarrow\infty)$. 
Thus, in this case neglecting $\rho_{ab}$  is not 
a good approximation as in the offresonant case. 
The results
reported in the main body of the paper correspond to the solution 
of the full Floquet-Markov 
equation, Eq.~\ref{eqFM}.
However, we have
verified that most of our results (including the dynamic transition discussed in Sec.III) are
accurately reproduced by the approximation that assumes
an asymptotic density matrix diagonal in the Floquet basis.

Assuming that the density matrix becomes diagonal in the Floquet basis, 
one can separate the dynamics of the diagonal and the off-diagonal density matrix.
The off-diagonal part is dominated by the dependence
$$ \frac{d\rho_{\alpha\beta}}{dt} \approx \left[-\frac{i}{\hbar}
(\varepsilon_\alpha-\varepsilon_\beta)+ L_{\alpha\beta\alpha\beta}\right]
\,\rho_{\alpha\beta}\;\;\;\; \alpha\not=\beta
$$
In this approximation, the decoherence rate between the $|\alpha\rangle$ and
the $|\beta\rangle$ Floquet state, 
is given by $\Gamma_{\alpha\beta}=-L_{\alpha\beta\alpha\beta}$.
The dynamics for the diagonal part of the density matrix gives
a rate equation for the population of the Floquet states
$P_\alpha=\rho_{\alpha\alpha}$:
\begin{eqnarray}
\frac{dP_\alpha}{dt}&=&\sum_{\beta} L_{\alpha\alpha\beta\beta} P_\beta\nonumber\\
&=& 2\sum_{\beta}R_{\alpha\alpha\beta\beta}P_\beta - R_{\beta\beta\alpha\alpha}P_\alpha
\end{eqnarray}
where $R_{\alpha\alpha\beta\beta}=\sum_n
g_{\alpha\beta}^n|A_{\alpha\beta}^n|^2$, after Eq.~(\ref{rates}).
It is simple to solve the above rate equation 
when we restrict to two levels.  With two Floquet states $|a\rangle,|b\rangle$, the asymptotic populations
are $P_a^\infty=R_{aabb}/(R_{aabb}+R_{bbaa})$, $P_b^\infty=1-P_a^\infty$;
and the  relaxation  rate   is
$\Gamma_r = 2(R_{aabb}+R_{bbaa})$. 
Using Eq.(\ref{rates}), we can
decompose the relaxation rate as a sum of  terms that
 describe virtual  n-photon transitions:\cite{grifonih}
 \begin{equation}
 \Gamma_r = \Gamma^{(0)} +\sum_{n\not=0} \Gamma^{(n)}\;,
 \end{equation}
with
\begin{equation}
\Gamma^{(n)}=2(g_{ab}^n|A_{ab}^n|^2+g_{ba}^n|A_{ba}^n|^2).
\end{equation}

\end{document}